# Age and genetic relationships among CB, CH and CR chondrites

Elias Wölfer[a,b*], Gerrit Budde[b,c], Thorsten Kleine[a,b]

[a]Max Planck Institute for Solar System Research, Justus-von-Liebig-Weg 3, 37077 Göttingen, Germany.

[b]Institut für Planetologie, University of Münster, Wilhelm-Klemm-Straße 10, 48149 Münster, Germany.

[c]Department of Earth, Environmental and Planetary Sciences, Brown University, Providence, RI 02912, USA.

[*]corresponding author: woelfer@mps.mpg.de

**Abstract**

The carbonaceous Bencubbin-like (CB), high-metal (CH), and Renazzo-like (CR) chondrites are metal-rich chondrites that have been suggested to be genetically linked and are sometimes grouped together as the CR chondrite clan. Of these, the CB and CH chondrites are thought to have formed in an impact-generated vapor-melt plume from material that may be isotopically akin to CR chondrites. We report Mo, Ti, Cr, and Hf-W isotopic data for CB and CH chondrites in order to determine their formation time, to assess whether these chondrites are genetically related, and to evaluate their potential link to CR chondrites. An internal Hf-W isochron for the CH chondrite Acfer 182 yields an age of 3.8±1.2 Ma after formation of Ca-Al-rich inclusions, which is indistinguishable from the mean Hf-W model age for CB metal of 3.8±1.3 Ma. The Mo isotopic data for CB and CH chondrites indicate that both contain some of the same metal and silicate components, which themselves are isotopically distinct. As such, the different Mo isotopic compositions of bulk CB and CH chondrites reflect their distinct metal-to-silicate ratios. CR metal exhibits the same Mo isotopic composition as CB and CH metal, but CR silicates have distinct Mo and Ti isotopic compositions compared to CB and CH silicates, indicating that CB/CH chondrites may be genetically related to CR metal, but not to CR silicates. Together, the new isotopic data are consistent with formation of CB and CH chondrites in different regions of a common impact-generated vapor-melt plume and suggest that the CB and CH metal may derive from a metal-rich precursor genetically linked to CR chondrites. The Hf-W systematics of CH and CB chondrites indicate that the impact occurred at 3.8±0.8 Ma after the formation of Ca-Al-rich inclusions and, hence, up to ~1 Ma earlier than previously inferred based on Pb-Pb chronology.

# 1 Introduction

Differences in the chemical and isotopic compositions of carbonaceous chondrites are thought to reflect varying abundances of their principal components, namely refractory inclusions, chondrules, matrix, and Fe-Ni metal (Alexander, 2019; Hellmann et al., 2023). These differences not only provide the basis for classifying carbonaceous chondrites into several distinct groups, but they also provide fundamental information about the processes operating before and during chondrite parent body accretion.

Among the carbonaceous chondrite groups, the carbonaceous Bencubbin-like (CB) and carbonaceous high-metal (CH) chondrites stand out by having extraordinarily high metal abundances, strong depletions in moderately volatile elements, and near-complete lack of primary matrix (e.g., Krot et al., 2002). They are also dominated by the presence of unusual petrological components, which are less common in other carbonaceous chondrites, such as non-porphyritic (e.g., cryptocrystalline) chondrules and compositionally zoned metal grains (e.g., Weisberg et al., 2001; Campbell et al., 2002, Krot et al., 2021). Like the carbonaceous Renazzo-like (CR) chondrites, the CB and CH chondrites show strong enrichments in $^{15}$N (Rubin et al., 2003; Van Kooten et al., 2016) and all three groups share similarities in O isotopes (i.e., they plot along a common O isotope mixing line distinct from that of other carbonaceous chondrites; Weisberg et al., 1995). This, together with the enhanced metal abundances of CR chondrites and general mineralogical and bulk chemical similarities between CR, CB, and CH chondrites has led to their classification as CR-clan meteorites (Weisberg et al., 1995; Krot et al., 2002). The CR-clan members also share the youngest formation ages among chondritic meteorites, with chondrule formation ages between ~3.6 Ma and ~4-5 Ma after the formation of CAIs for CR and CB chondrites, respectively (Krot et al., 2005; Bollard et al., 2015; Schrader et al., 2017, Budde et al., 2018).

The unusual properties of the CB and CH chondrites have led to the proposal that these samples formed in a vapor-melt plume resulting from a collision between planetesimals (Wasson and Kallemeyn, 1990; Krot et al., 2005). Subsequent studies have suggested that these planetesimals were chemically differentiated, and, given the many similarities between CB, CH, and CR chondrites, that at least one of the colliding bodies was a CR chondrite-like object (e.g., Fedkin et al., 2015; Oulton et al., 2016; Koefoed et al., 2022). In this model, CB chondrites consist almost entirely of metal and silicate condensates from the impact-generated vapor plume (e.g., non-porphyritic chondrules and newly condensed metal grains; Oulton et al., 2016; Weyrauch et al., 2019; Krot et al., 2021), while CH chondrites may have formed in more distal regions of the vapor plume and also incorporated a larger share of solar nebula material from outside the plume [e.g., porphyritic chondrules, chondritic lithic clasts, CAIs surrounded by Wark-Lovering rims; (Krot et al., 2021; Weyrauch et al., 2021)]. Thus, if CB and CH chondrites formed from the same vapor plume, then they should contain the same metal and silicate components, which were mixed in different proportions to account for the higher metal abundance in CB than CH chondrites.

Meteorites that are genetically linked should be characterized by the same nucleosynthetic isotope composition. Nucleosynthetic isotope anomalies are mass-independent deviations from the terrestrial composition and arise through the heterogeneous distribution of presolar material in the solar nebula. In particular, these anomalies allow distinguishing between non-carbonaceous (NC) and carbonaceous (CC) meteorites, which represent two spatially distinct but contemporaneous dust reservoirs of the solar protoplanetary disk (Warren, 2011; Budde et al., 2016a). There are additional isotope variations within each reservoir, which make it possible to establish genetic links among the members of each group (e.g., Kleine et al., 2020; Kruijer et al., 2022) and to assess how the heterogeneous distribution of chondrite components affects the isotopic variations among bulk chondrites (e.g., Alexander, 2019; Hellmann et al., 2023).

For instance, CR, CB, and CH chondrites have indistinguishable $^{54}$Cr isotopic signatures typical for CC meteorites, consistent with the idea that these three chondrite groups are genetically linked (Zhu et al., 2021). However, bulk samples of CR, CB, and CH chondrites are characterized by distinct Mo isotopic compositions (Budde et al., 2019), indicating that these three chondrite groups did not simply form from the same bulk material. These results suggest that a more systematic study involving isotope signatures of lithophile and siderophile elements as well as component-specific analyses are needed to better characterize the genetic relationships among silicates and metal in the CR-clan meteorites.

Formation of CB and CH chondrites from the same impact-generated vapor-melt plume implies that both have identical formation ages. However, unlike for CB chondrites, which have been dated by several chronometers to have formed at ~4-5 Ma after CAI formation (e.g., Krot et al., 2005; Bollard et al., 2015; Pravdivtseva et al., 2017), there are no precise ages for CH chondrites. The short lived $^{182}$Hf-$^{182}$W chronometer ($t_{1/2} \approx 8.9$ Ma; Vockenhuber et al., 2004) is well suited to date the formation of metal-bearing chondrites (e.g., Budde et al., 2018; Hellmann et al., 2019), but an earlier attempt to date CH chondrites using the Hf-W system in metal and silicate samples provided only an imprecise age, reflecting the limited spread in Hf-W ratios among the samples and the lower precision of W isotope measurements achievable at the time of this earlier study (Kleine et al., 2005). Tungsten isotopes can now be measured with much higher precision (e.g., Kruijer et al., 2014; Kruijer and Kleine, 2018), making it theoretically possible to date CH chondrites more precisely.

The objectives of this study are to better constrain the genetic and age relationships among CR-clan chondrites using their Mo, Ti, and Cr, and Hf-W isotope systematics. These data will be used to further our understanding of the formation of CB and CH chondrites within the framework of existing formation models that invoke condensation in an impact-generated vapor-melt plume, and to shed new light on the genetic link between CB/CH and CR chondrites.

## 2 Materials and Methods

*2.1 Sample preparation*

The metal-rich chondrite samples investigated in the present study include the CH3 chondrite Acfer 182, the CH/CB$_b$ chondrite breccia Isheyevo, three CB chondrites (Hammadah al Hamra 237 (HaH 237), Gujba, Bencubbin), and the CR2 chondrite Acfer 139. Of these, Acfer 182 is paired with Acfer 207 and Acfer 214 (Bischoff et al., 1993b), Acfer 139 is paired with eight other Acfer meteorites (059, 087, 097, 114, 186, 187, 209, 270) as well as El Djouf 001 (Bischoff et al., 1993a), and HaH 237 displays chemical and petrologic properties similar to Queen Alexandra Range 94411 and 94627 (QUE 94411 and QUE 94627; Krot et al., 2001; Campbell et al., 2005). Based on its very high metal content (~70 vol.%), the occurrence of zoned metal grains, and its small average grain size, HaH 237 is classified as a CB$_b$ chondrite (Weisberg et al., 1995; Krot et al., 2002). By contrast, Gujba and Bencubbin are CB$_a$ chondrites, as they contain less and compositionally uniform metal nodules (~40−60 vol.%), and are texturally coarse-grained (up to 0.5−1 cm-sized metal grains; Weisberg et al., 1995, 2001).

The CB and CH chondrites of this study are of petrological type 3.00 and are among the most pristine materials formed in the early Solar System (Bischoff et al., 1993a, b; Ivanova et al., 2008). By contrast, Acfer 139 is of petrologic type 2, which is typical for CR chondrites. Acfer 182, Bencubbin, and most lithologies of Gujba and Isheyevo display only low shock degrees (S1–S2), whereas HaH 237 and individual lithologies of Isheyevo show petrologic evidence for substantial thermal processing and shock metamorphism of stage S3−S4 (Ivanova et al., 2008) or, in the case of Gujba, even show metal-silicate impact melted regions (Krot et al., 2002).

Metal and silicate components of Acfer 182 were manually separated following the procedure described by Budde et al. (2018). Three adjacent pieces (~4.5 g in total) of Acfer 182 were polished with SiC and sonicated in ethanol to remove adhering dust particles. Afterwards, the pieces were gradually crushed in an agate mortar and sieved through polyamide meshes. During grinding, coarse metal grains were removed using a hand magnet or by hand-picking, and collected as coarse metal fraction 'M-1' (>125 μm). The fine-grained rock powder (<40 μm) after the last sieving step (125 μm) was collected as fines fraction 'F-1' (~0.55 g). The remaining part of the sample, consisting of metal and silicate grains between 40-125 μm, was washed with ultrapure water and ethanol to remove adhering dust, and then further separated. A weak hand magnet was used to remove pure metal grains, which were collected as fine metal fraction 'M-2' (40-125 μm). The residual material was split into five fractions according to the magnetic susceptibilities using different hand magnets. These fractions were termed 'MS-$n$' ($n$ = 1, 2, 3, 4, 5), which consist of variable amounts of metal, matrix, chondrules, and other silicate phases. MS-1 displays the most magnetic mixed fraction (i.e., enriched in components with high magnetic susceptibility), whereas MS-5 is composed of the least magnetic components. Each of the five mixed fractions consisted of 0.40–0.55 g material.

Metal grains in CH chondrites are often intergrown with silicates. To obtain metal separates as pure as possible, the metal fractions M-1 and M-2 were ground in an agate mortar filled with ethanol. The silicates were removed from the metal by mechanical spalling and then collected in suspension. After discarding the supernatant, the remaining material was dried and the procedure repeated until the supernatant became clear. By this approach, ~0.15 g and ~0.30 g of pure metal was obtained for the fractions M-1 and M-2, respectively. Additionally, metal separates were obtained for Gujba (CB$_a$), Bencubbin (CB$_a$) and Isheyevo (CB$_b$/CH), where the sample preparation was the same as for the metal separates of Acfer 182. Finally, whole rock data for Acfer 182 (CH), HaH 237 (CB$_b$), Acfer 139 (CR2), and Graves Nunataks 06100 (GRA 06100; CR2.5) were obtained from ~0.5 g pieces that were ground to fine bulk powders.

*2.2   Chemical separation and isotope measurements*

Sample digestion protocols were broadly adapted from Budde et al. (2018). Whole-rock powders and MS-*n* mixed fractions were digested via table-top digestion using HF-HNO$_3$-HClO$_4$ (2:1:0.01) at 180-200 °C (5 days), followed by 'inverse' aqua regia (2:1 HNO$_3$-HCl) at 130-150 °C (2 days). To reduce potential W contamination from the digestion beaker, the least magnetic mixed fraction MS-5 was digested in HF-HNO$_3$ (2:1) only at 130 °C (4 days), followed by inverse aqua regia at 110 °C (1 day). The metal fractions M-1 and M-2 were dissolved in inverse aqua regia (+ trace HF) at 110-130 °C (2 days). After digestion, all samples were dissolved in 40 ml 6 M HCl – 0.06 M HF, and small aliquots were taken for isotope dilution analysis to determine Hf and W concentrations.

The chemical separation and isotope measurements of W, Mo, and Ti followed the established analytical protocols of our laboratory (see Budde et al. (2018) and Gerber et al. (2017) for details). Isotope measurements were performed on the Thermo Scientific® Neptune Plus MC-ICP-MS in the Institut für Planetologie at the University of Münster. Tungsten was separated from the sample matrix using a two-stage anion exchange chromatography, where the yields were typically ~70% and total procedural blanks were negligible (<150 pg W). For the W isotope analyses, instrumental mass bias was corrected by internal normalization to $^{186}$W/$^{184}$W = 0.92767 ('6/4') or $^{186}$W/$^{183}$W = 1.98590 ('6/3') using the exponential law. Isobaric (Os) interference corrections were negligible for all analyzed samples. The W isotope compositions of samples were measured relative to an Alfa Aesar® solution standard and are reported as ε-unit deviations (*i.e.*, 0.01%) relative to the mean of the bracketing standard analyses. For samples analyzed multiple times, the reported values represent the mean of pooled solution replicates. The accuracy and precision of the W isotope measurements were assessed by repeated analyses of the SRM 129c and BHVO-2 reference materials, which show indistinguishable W isotope compositions and define an external reproducibility (2 s.d.) of ~0.1 ε for all W isotope ratios. Tungsten isotope ratios that involve $^{183}$W show a small analytical artifact (~0.1 ε) for the terrestrial standards (see Budde et al., 2022); after correction for this 'analytical $^{183}$W effect', all isotope ratios are indistinguishable from the Alfa Aesar® standard (**Table S1**). For all investigated samples, $\varepsilon^{i}$W values that involve $^{183}$W were corrected by using the mean values obtained for the terrestrial standards (see Kruijer et al., 2014), and the associated uncertainties induced through this correction were propagated into the final uncertainties reported for the W data.

Molybdenum was collected from column washes at the end of both steps of the two-stage anion exchange chemistry used for the separation of W. The 'Mo cuts' were combined and Mo concentrations were determined using a Thermo Scientific® XSeries 2 quadrupole ICP-MS in Münster. Molybdenum was further purified using a two-stage ion exchange chromatography (Budde et al., 2018), where typical yields were ~75% and total procedural blanks were negligible (~2–4 ng Mo). For the Mo isotope analyses, instrumental mass bias was corrected by internal normalization to $^{98}$Mo/$^{96}$Mo = 1.453173 using the exponential law. Isobaric interferences of Zr (on $\varepsilon^{94}$Mo) and Ru (on $\varepsilon^{100}$Mo) were corrected by monitoring interference-free $^{91}$Zr and $^{99}$Ru, respectively. All final Mo cuts had Ru/Mo and Zr/Mo of <1×10$^{-4}$, and thus were well within the correctable range (Budde et al., 2016a). As for W, Mo isotope compositions were measured relative to the mean of bracketing runs of anAlfa Aesar® solution standard and are reported as ε-unit deviations from the terrestrial standard. For samples analyzed multiple times, the reported values represent the mean of pooled solution replicates. The accuracy and precision of the Mo isotope measurements were assessed by repeated analyses of the BHVO-2 standard [data reported in Budde et al. (2019)]. The $\varepsilon^{i}$Mo values obtained for BHVO-2 are indistinguishable from the Alfa Aesar® standard, demonstrating that the Mo isotopic data are accurate. The external reproducibility (2 s.d.) of the Mo isotope measurements ranges from ±0.15 for $\varepsilon^{97}$Mo to ±0.35 for $\varepsilon^{92}$Mo.

Titanium was collected during the clean-up step of the two-stage anion exchange chromatography used for the separation of W, where Ti is eluted in 1 M HCl–2 wt.% $H_2O_2$. Titanium concentrations in these column washes were determined using a Thermo Scientific® XSeries 2 quadrupole ICP-MS in Münster and aliquots (equivalent to ~30 µg Ti) were further purified using a two-stage anion exchange chromatography. The yields of this clean-up procedure were typically 90–95% and the total procedural blanks were negligible (~2-3 ng Ti). Titanium isotope compositions were measured in high resolution mode and instrumental mass bias was corrected by internal normalization to $^{49}Ti/^{47}Ti = 0.749766$ using the exponential law. Isobaric interferences of Ca (on $\varepsilon^{46}Ti$ and $\varepsilon^{48}Ti$), V (on $\varepsilon^{50}Ti$), and Cr (on $\varepsilon^{50}Ti$) were corrected by monitoring interference-free $^{44}Ca$, $^{51}V$, and $^{53}Cr$, respectively. The final Ti cuts had Ca/Ti of <5×10$^{-3}$, and thus were well within the correctable range (Zhang et al., 2011). Isobaric V and Cr interference corrections were negligible for all analyzed samples. The Ti isotope data are reported as ε-unit deviations relative to bracketing measurements of the Origins Lab OL-Ti solution standard. The reported values represent the mean of pooled solution replicates.. The accuracy and precision of the Ti isotope measurements were assessed by frequent repeated analyses of the JB-2 standard, which show indistinguishable Ti isotope compositions compared to the OL-Ti standard and define an external reproducibility (2 s.d.) of ±0.14 for $\varepsilon^{48}Ti$ and ±0.27 for $\varepsilon^{50}Ti$ (**Table S2**).

In addition, for one sample of this study, Acfer 182 (CH), we also report Cr isotope data. The methods for Cr purification by ion exchange chromatography and isotope measurements by thermal ionization mass spectrometry followed our established protocols (Schneider et al., 2020).

## 3 Results

### 3.1 Chemical composition

Concentrations of Ti, Cr, Hf, Mo, and W for the CH, CB, and CR chondrites are provided in **Table 1** and **2**. As expected, these concentrations vary systematically with the metal content of the samples. The highest Mo and W concentrations are observed for the metal fractions of all samples. These are generally characterized by low $^{180}$Hf/$^{184}$W <0.035 (except for Isheyevo; $^{180}$Hf/$^{184}$W = 0.065), indicating that relatively pure metal separates were obtained. For the CH chondrites Acfer 182, coarse-grained metal has higher W and Mo concentrations than fine-grained metal, suggesting that siderophile element concentrations in CH metals are grain-size-dependent, as has previously been observed for CR chondrite metal (Budde et al., 2018). Bulk samples of Acfer 139 (CR), Acfer 182 (CH), and HaH 237 (CB) display decreasing Ti and Hf and increasing Mo and W concentrations, consistent with the different abundances of Fe-Ni metal in these chondrites (CB > CH > CR). As a result, the $^{180}$Hf/$^{184}$W ratios of ~1.3, ~0.9, and ~0.2 for these samples decrease from the typical chondritic value observed for CR chondrites to lower values in the more metal-rich CH and CB chondrites. The metal-silicate mixed fractions (MS-*n*) of Acfer 182 show variable Ti, Hf, Mo, and W concentrations, consistent with the increasing amounts of Mo,W-rich metal and decreasing amounts of Ti,Hf-rich silicates with increasing magnetic susceptibility of these fractions (**Fig. 1**). As a result, the MS-*n* fractions display variable $^{180}$Hf/$^{184}$W ratios ranging from ~0.4 for MS-1 to ~2.8 for the least magnetic fraction MS-5.

### 3.2 Mo, Ti, and Cr isotopes

As expected from their classification as carbonaceous chondrites, all samples of this study plot on or close to the carbonaceous chondrite (CC)-line in $\varepsilon^{94}$Mo vs. $\varepsilon^{95}$Mo space (**Fig. 2**,

**Table 1**). In this diagram, meteorites plot along two approximately parallel *s*-process mixing lines, where the offset between these two lines probably reflects an *r*-process excess in the CC over the NC reservoir (e.g., Budde et al., 2016a).

The CB and CH metals exhibit indistinguishable Mo isotopic compositions (mean $\varepsilon^{94}$Mo = 1.29 ± 0.05; 95% CI, *n* = 5), which are also indistinguishable from the average Mo isotopic composition of CR metal (Budde et al., 2018). Bulk samples of CB (HaH 237), CH (Acfer 182), and CR chondrites (Acfer 139) display different $\varepsilon^{94}$Mo values of ~1.3, ~1.8, and ~2.7, respectively, indicating that the $\varepsilon^{94}$Mo values increase systematically with decreasing metal content of the samples. Similar to the bulk meteorites, the metal-silicate mixed fractions of Acfer 182 also display increasing $\varepsilon^{94}$Mo (e.g., from ~1.5 to ~2.5) with decreasing magnetic susceptibility. Overall, the $\varepsilon^{94}$Mo values are well-correlated with 1/Mo and $^{180}$Hf/$^{184}$W (**Fig. 3**), indicating that the metal-silicate ratio exerts a strong control on the Mo isotope signatures of the CB, CH and CR chondrites.

All samples of this study exhibit well-resolved $\varepsilon^{50}$Ti and $\varepsilon^{46}$Ti anomalies (**Table 1**) and plot along the $\varepsilon^{50}$Ti–$\varepsilon^{46}$Ti correlation line (**Fig. 4**) defined in prior studies (e.g., Trinquier et al., 2009; Torrano et al., 2019). Unlike for Mo and W, the whole-rock sample and all magnetic separates of Acfer 182 (CH) display homogeneous $\varepsilon^{46}$Ti and $\varepsilon^{50}$Ti excesses of 0.35 ± 0.04 and 1.97 ± 0.06, respectively (95% CI, *n* = 7; including F-1), which are indistinguishable from the Ti isotopic composition of the bulk CB chondrite of this study (**Fig. 4**). By contrast, the $\varepsilon^{46}$Ti and $\varepsilon^{50}$Ti anomalies of bulk CR chondrites (Acfer 139 (this study); Northwest Africa 801 (NWA 801; Zhang et al., 2012); **Table 1**) are 0.46 ± 0.04 and 2.43 ± 0.21 (2σ, *n* = 2), respectively, and thus resolved from the values of bulk CH and CB samples.

Finally, the $\varepsilon^{54}$Cr value of 1.41 + 0.13 measured for the CH chondrite Acfer 182 determined in this study is in good agreement with results of prior studies (Trinquier et al., 2007; Zhu et al., 2021).

*3.3 Hf-W systematics*

As expected from their variable $^{180}$Hf/$^{184}$W ratios, the samples of this study exhibit $^{182}$W variations, with ε$^{182}$W (6/3) values ranging from approximately –3.1 to –0.6 (**Table 2**). Of note, all CB, CH and CR metals have indistinguishable ε$^{182}$W values of around –3.1. All other samples have more radiogenic $^{182}$W compositions, which are correlated with their $^{180}$Hf/$^{184}$W ratios (**Fig. 5**). This includes the different magnetic separates of Acfer 182 (MS-1, MS-2) as well as the bulk samples of Acfer 139 (CR), Acfer 182 (CH), and HaH 237 (CB).

All samples of this study show $^{183}$W excesses (**Fig. 6**), indicative of nucleosynthetic W isotope anomalies. For instance, the combined fractions of Acfer 182 are characterized by a mean ε$^{183}$W of 0.15 ± 0.04 [95% confidence interval (95% CI), *n* = 9], the combined CB samples by ε$^{183}$W = 0.11 ± 0.03 [without Isheyevo; 2 s.d., *n* = 3], and the CB and CH metal separates by ε$^{183}$W = 0.12 ± 0.03 [95% CI, *n* = 5], which is indistinguishable from the average ε$^{183}$W of CR metal (Budde et al., 2018). A bulk sample of the CR chondrite Acfer 139 exhibits a larger ε$^{183}$W excess of 0.45 ± 0.16, consistent with previously published data for bulk CR chondrites (Budde et al., 2018). The presence of small nucleosynthetic W isotope anomalies is also evident from the difference in measured ε$^{182}$W (6/3) and ε$^{182}$W (6/4) values. Due to the varying contributions of the *s*-process to individual W isotopes, and because $^{184}$W has the largest *s*-process contribution, the effects of nucleosynthetic heterogeneity are largest for ε$^{182}$W (6/4) and minor to absent for ε$^{182}$W (6/3) (see Section **5** for details). This is consistent with the observation that the largest ε$^{182}$W (6/3)–ε$^{182}$W (6/4) offset is observed for the sample with the largest $^{183}$W anomaly (Acfer 139).

## 4 Genetic links among and between CB, CH and CR chondrites

### 4.1 Common heritage of CB and CH chondrites

Nucleosynthetic isotope anomalies can be used to assess genetic relationships among meteorites and their components, because only materials with the same nucleosynthetic isotope signatures are genetically related. Recently, Zhu et al. (2021) measured the Cr isotopic composition of two bulk CH and three bulk CB chondrites and found indistinguishable $\varepsilon^{54}$Cr values of ~1.5, which is consistent with $\varepsilon^{54}$Cr = 1.41±0.13 measured for the CH chondrite Acfer 182 in this study, and also with Cr isotopic data previously reported for CB and CH chondrites by Trinquier et al. (2007) and Yamashita et al. (2010). Similarly, nucleosynthetic Ti isotope data for bulk CB and CH chondrites of this study are indistinguishable and show a common $^{50}$Ti excess of $\varepsilon^{50}$Ti ≈ 2, in line with Ti isotopic data reported for CB chondrites by Trinquier et al. (2009). Thus, the Cr and Ti isotopic data suggest that silicate phases in CB and CH chondrites are co-genetic, which is consistent with a more general genetic link inferred between these two chondrite groups based on their N and O isotope characteristics (Krot et al., 2002, 2012; Van Kooten et al., 2016).

Information on potential genetic relationships among the metal of CB and CH chondrites can be obtained from nucleosynthetic isotope anomalies in siderophile elements, such as Ru, W, and Mo. For instance, further support for a close genetic link between CB and CH chondrites comes from their indistinguishable nucleosynthetic Ru (Fischer-Gödde and Kleine, 2017) and W (this study) isotope signatures. However, for these two elements nucleosynthetic isotope variations among different carbonaceous chondrites are not systematically resolved and, therefore, do not allow to firmly establish genetic links between two specific groups of carbonaceous chondrites. More stringent constraints on the genetics of CB and CH metals are provided by their Mo isotope signatures determined in this study. Contrary to the similarities discussed so far, bulk CB and CH chondrites display distinct Mo isotope signatures, yet metal from both chondrites has indistinguishable Mo isotopic compositions (e.g., $\varepsilon^{94}$Mo ≈ 1.3). Importantly, the mixed metal-silicate separates of Acfer 182 plot along a linear trend in a diagram of $\varepsilon^{94}$Mo versus 1/Mo, which passes through the isotopic composition of the CH and CB metals (**Fig. 3**). The same is observed when $\varepsilon^{94}$Mo is plotted against $^{180}$Hf/$^{184}$W. Since the variable $^{180}$Hf/$^{184}$W ratios are indicative of different silicate-to-metal ratios, this linear trend demonstrates that the CH and CB meteorites can be described as mixtures of the same metal and silicate components, which themselves are isotopically distinct from one another. Within this framework, CB chondrites are (consistent with petrographic observations) enriched in the metal component over the CH chondrites. The distinct Mo isotope signatures of *bulk* CH and CB chondrites are, therefore, supporting a genetic link as suggested by the other isotope systems, as they merely reflect varying degrees of mixing between the same metal phase (e.g., $\varepsilon^{94}$Mo ≈ 1.3) and the same isotopically distinct silicate phase, which is characterized by larger Mo isotope anomalies.

All metal-silicate fractions of Acfer 182 plot along the CC-line in the $\varepsilon^{94}$Mo–$\varepsilon^{95}$Mo diagram (**Fig. 2**), indicating that the Mo isotopic difference between CH metal and silicates are governed by *s*-process variations, where the silicates have a larger *s*-process deficit than the metal. A similar observation has been made previously for metal and silicates from CR chondrites, where the silicates are also *s*-process-depleted compared to the metal (Budde et al., 2018), and for matrix and chondrules from CV chondrites, where chondrules are depleted in *s*-process Mo nuclides relative to the matrix (Budde et al., 2016a). For the CR and CV chondrites this has been interpreted to indicate transfer of an *s*-process-enriched, metallic carrier from the silicates/chondrules into the metal/matrix, most likely during chondrule formation (Budde et al., 2016a, b, 2018). For the CV chondrites it has also been suggested that the *s*-process-enriched nature of matrix over chondrules reflects transference of *s*-process-depleted Mo from the matrix into chondrules during parent body metamorphism (Sanders and Scott, 2022). This model, however, has difficulties to account for the observation that in CR chondrites, metal is *s*-process-enriched compared to co-existing silicates, indicating that the *s*-process-enriched signatures are indeed carried by metal (Budde et al., 2018). Moreover, the data of this study show that CH chondrites show the same systematic *s*-process Mo isotopic difference between metal and silicates, yet these samples largely escaped modifications by parent body processes such as thermal metamorphism and aqueous alteration (e.g., Bischoff et al., 1993a, b; Krot et al., 2002). As such, any redistribution of *s*-process Mo on the CH chondrite parent body is highly unlikely, and so the distinct Mo isotopic signatures of CH chondrite metal and silicates were almost certainly established during formation of the CH chondrites. Given the similar pattern of *s*-process Mo isotope heterogeneity in CH, CR, and CV chondrites (i.e., *s*-process-enrichment in metal/matrix over silicates), it seems likely that these heterogeneities were established by similar processes, such as transfer of a metallic *s*-process carrier from the chondrules into the metal or matrix.

In summary, the common Mo isotope signatures of CB and CH metals, combined with the Mo isotope variations of mixed metal-silicate fractions of CH chondrites, indicate that CB and CH chondrites contain some of the same precursor materials. In particular, these data show that Mo isotopic differences between bulk samples of CB and CH chondrites correlate with their different metal-to-silicate ratios, suggesting strongly that the dominant metal and silicates in CB and CH chondrites share the same genetic heritage. As such, the larger fraction of solar nebula materials in CH compared to CB chondrites does not seem to have a measurable effect on the isotopic composition of bulk CH chondrites.

*4.2   Genetic difference between CR and CB-CH chondrites*

As noted above, CR chondrites have been associated with CB and CH chondrites and grouped together as metal-rich carbonaceous chondrites in the CR-clan (e.g., Weisberg et al., 2001; Krot et al., 2002). The isotopic data of this study can be used to shed new light on potential genetic links between CR chondrites and the metal and silicates of the CB and CH chondrites.

Metal from CR, CB, and CH chondrites have indistinguishable Mo isotopic compositions, suggesting a genetic link between these metals. Moreover, in the $\varepsilon^{94}$Mo–$\varepsilon^{95}$Mo diagram, bulk CR chondrites plot on the CC-line towards larger *s*-process Mo isotope deficits relative to CB and CH chondrites (**Fig. 2**), consistent with their higher proportion of (*s*-process-depleted) silicates. Based on these observations, it is tempting to regard CR chondrites as mixtures of the very same two, isotopically distinct metal and silicate components that also represent the building material of CB and CH chondrites. However, CR chondrites do not follow the same trend as CB and CH chondrites in the $\varepsilon^{94}$Mo versus 1/Mo (or $^{180}$Hf/$^{184}$W) plot, indicating that CR silicates have a different Mo isotopic composition than and, hence, are genetically unrelated to silicates in CB and CH chondrites (**Fig. 3**). This is also evident from the distinct Ti isotopic compositions of CR ($\varepsilon^{50}$Ti ≈ 2.5) and CB/CH chondrites ($\varepsilon^{50}$Ti ≈ 2); as a lithophile element, Ti records the genetic heritage of solely the silicates of these chondrites, such that these different $\varepsilon^{50}$Ti values indicate a distinct heritage of silicates in CR versus CB and CH chondrites. By contrast, all three chondrite groups share a common $\varepsilon^{54}$Cr composition, which also records the heritage of silicates in these chondrites. Prior studies have shown that the abundance of Ca-Al-rich inclusions (CAI) exerts a strong control on the Ti isotopic composition of carbonaceous chondrites without affecting their Cr isotopic composition, reflecting the high Ti/Cr ratio and strong Ti enrichment of CAIs (e.g., Trinquier et al., 2009; Torrano et al., 2021; Hellmann et al., 2023). Thus, one possibility is that the higher $\varepsilon^{50}$Ti of CR chondrites reflects a larger proportion of CAIs in these chondrites compared to the CB and CH chondrites, while the remaining silicates share a common origin.

In summary, while the CR, CB, and CH chondrites contain metal with indistinguishable Mo isotopic compositions, suggesting a common heritage of these metals, silicates in CR chondrites have distinct Mo and Ti isotopic compositions compared to silicates in CB and CH chondrites. Thus, although all three chondrite groups share many similarities, at least some of their building material appears to have a different origin, indicating an important genetic difference between CR and CB/CH chondrites.

## 5  Hf-W chronology of CB and CH chondrites

A pre-requisite for application of the Hf-W system to date CB and CH chondrites is that W in the components of these samples was once in isotopic equilibrium. However, the presence of nucleosynthetic W isotope variations between metal and silicates in these samples reveal that this was not entirely the case, raising the question of whether these samples can nevertheless be dated using the Hf-W system. Several lines of evidence indicate that this is the case. First, the nucleosynthetic W isotope variations among the constituents of carbonaceous chondrites are likely governed by the heterogeneous distribution of isotopically highly anomalous presolar carrier phases, which account for only a negligible amount of the total W budget of a sample. Thus, despite the preservation of nucleosynthetic W isotope variations, nearly the total W budget may have nevertheless been in isotopic equilibrium (Budde et al., 2016b, 2018). Second, for $\varepsilon^{182}$W (6/3) nucleosynthetic isotope variations are very small (<0.02 $\varepsilon$ for the samples of this study), because nucleosynthetic W isotope anomalies predominantly reside on $^{184}$W, which is not used in this normalization (e.g., Burkhardt et al., 2008; Kruijer et al., 2014; Budde et al., 2016b). Thus, the observed $\varepsilon^{182}$W (6/3) variations can entirely be attributed to radiogenic variations; this is consistent with the observation that for the samples of this study, measured ε¹⁸²W (6/3) values correlated with the $^{180}$Hf/$^{184}$W ratios of the samples, as expected for purely radiogenic variations among samples having the same age. Finally, as shown below, the Hf-W systematics of metal and silicates from the CH chondrite Acfer 182 cannot easily accounted for by simple metal-silicate mixing, but instead indicate identical formation ages for metal and silicates in this sample.

*5.1 Model age for CB metal*

CB chondrites contain only small amounts of silicates (e.g., Weisberg et al., 2001), and so constructing a Hf-W isochron based on metal and silicate separates is not straightforward. However, Hf-W data for metal-rich samples can be used to calculate Hf-W model ages for metal-silicate separation by assuming that the metal formed by a single event of metal segregation from a reservoir having a chondritic Hf/W ratio. The model age is calculated as follows (e.g., Horan et al., 1998):

$$\Delta t_{CAI} = -\frac{1}{\lambda} \times \ln\left[\frac{\varepsilon^{182}W_{metal} - \varepsilon^{182}W_{CHUR}}{\varepsilon^{182}W_{SSI} - \varepsilon^{182}W_{CHUR}}\right]$$

where $\varepsilon^{182}W_{metal}$ is the W isotope composition of the metal, $\varepsilon^{182}W_{CHUR} = -1.9 \pm 0.1$ is the present-day W isotope composition of chondrites (Kleine et al., 2002, 2004; Yin et al., 2002; Schoenberg et al., 2002), $\varepsilon^{182}W_{SSI} = -3.49 \pm 0.07$ defines the Solar System initial W isotope composition as inferred from CAIs (Kruijer et al., 2014a), and $\lambda = 0.0778 \pm 0.0015$ Ma$^{-1}$ is the decay constant of $^{182}$Hf (Vockenhuber et al., 2004).

All CB metal separates of this study contain some Hf, and so their measured ε¹⁸²W must first be corrected for radiogenic ingrowth from $^{182}$Hf decay using the measured $^{180}$Hf/$^{184}$W of a sample and its initial $^{182}$Hf/$^{180}$Hf. The latter is not known a priori (because the age of the sample is not known), but can be inferred iteratively. For this a model age is calculated using a sample's measured $\varepsilon^{182}$W, and this age is used to infer the sample's initial $^{182}$Hf/$^{180}$Hf. This is used to correct the measured $\varepsilon^{182}$W for radiogenic ingrowth (correction is <0.03$\varepsilon$), which is then used to calculate a new model age and hence initial $^{182}$Hf/$^{180}$Hf. This procedure is repeated until the inferred initial $^{182}$Hf/$^{180}$H no longer changes (Budde et al., 2015).

The model ages for the two CB metals of this study are indistinguishable from each other and are 3.3±2.0 and 4.2±1.8 Ma after CAI formation, respectively, with an average model age for the combined CB metals of 3.8±1.3 Ma [2 s.d., *n* = 2; **Table S3**; **Fig. 7**] after CAI formation. Metal separates from the CH/CB$_b$ chondrite breccia Isheyevo display a similar Hf-W model age of 3.7±1.7 Ma after CAI formation (**Fig. 7**). Consistent with the 'late' Hf-W model age for CB metal, Al-Mg data for CB$_a$ and CB$_b$ chondrules reveal no radiogenic $^{26}$Mg excesses, indicating they formed later than ~3.5 Ma after CAI formation (Gounelle et al., 2007; Olsen et al., 2013; Nagashima et al., 2018). The Hf-W age is also in good agreement with a Mn-Cr age of 4.0±0.9 Ma after CAI formation obtained from a Mn-Cr isochron for chondrules and metal grains from Gujba (Yamashita et al., 2010) [re-calculated using an initial $^{53}$Mn/$^{55}$Mn of (3.24 ± 0.04) × 10$^{-6}$ (Glavin et al., 2004) and an U-corrected Pb-Pb age of 4563.37 ± 0.25 (Brennecka and Wadhwa, 2012) for the D'Orbigny angrite, and a CAI formation age of 4567.3 ± 0.2 (Connelly et al., 2012)], and with an I-Xe age of Gujba chondrules of 4.1±1.3 Ma after CAI formation (Gilmour et al., 2009) [using the Shallowater aubrite with an initial $^{129}$I/$^{127}$I of (1.072 ± 0.02) × 10$^{-4}$ (Brazzle et al., 1999) and an U-corrected absolute Pb-Pb age of 4562.3 ± 0.8 (Gilmour et al., 2009)]. Pravdivtseva et al. (2017) reported an I-Xe age for a chondrule from HaH 237 of 5.2±0.3 Ma after CAI formation [relative to the 4562.4 ± 0.2 Ma age for Shallowater], which seems to be slightly younger than albeit not resolved from the Hf-W model age, and also the aforementioned Mn-Cr and I-Xe ages for Gujba (**Fig. 8**).

In contrast to this good agreement among ages, the Hf-W ages tend to be slightly older than the Pb-Pb ages for CB chondrites. Bollard et al. (2015) reported a weighted mean Pb-Pb age of 4562.5±0.2 Ma (corrected for U isotope variations using $^{238}$U/$^{235}$U = 137.786 ± 0.013) for individual chondrules of Gujba, which translates into a relative age of 4.8±0.3 Ma after CAI formation using a Pb-Pb age for CAIs of 4567.3±0.2 Ma (Connelly et al., 2012). Krot et al. (2005), however, reported somewhat younger Pb-Pb ages for chondrules from HaH 237 and Gujba of 5.5±0.9 Ma and 5.6±0.5 Ma after CAI formation, respectively [note that these ages have been re-calculated using $^{238}$U/$^{235}$U = 137.786 ± 0.013; see Bollard et al. (2015)]. Some additional uncertainty in these ages arises because the exact value for the Pb-Pb age of CAI is debated, and older ages of up to 4568.2±0.2 Ma have been proposed as the more accurate formation ages of CAIs (Bouvier and Wadhwa, 2010; Piralla et al., 2023). Using this older CAI age instead of the 4567.3 Ma age used above would result in relative Pb-Pb ages of CB chondrites that are ~1 Ma younger. Either way, the Pb-Pb ages for CB chondrites seem to be slightly younger than their Hf-W age, although this difference is not well resolved (**Fig. 8**). We note, however, that the Pb-Pb ages for individual Gujba chondrules appear to be resolvably different, suggesting that the Pb-Pb system in these samples does not date a single event of CB chondrite formation, but rather that individual samples may have at least partially been reset. We, therefore, interpret the Hf-W model age of CB metals of 3.8±1.3 Ma to date metal-silicate fractionation during the original formation of these metals.

*5.2   Internal isochron for Acfer 182*

The different fractions of the CH chondrite Acfer 182 define a linear correlation in a plot of $\varepsilon^{182}$W versus $^{180}$Hf/$^{184}$W that, if interpreted as an isochron, defines an initial $^{182}$Hf/$^{180}$Hf = (7.60 ± 0.60) × 10$^{-5}$ (MSWD = 0.17, *n* = 9) and an initial $\varepsilon^{182}$W = –3.10 ± 0.09 (**Fig. 5a**). Relative to the ($^{182}$Hf/$^{180}$Hf)$_i$ = (1.018 ± 0.043) × 10$^{-4}$ of CAIs (Kruijer et al., 2014a), this initial $^{182}$Hf/$^{180}$Hf yields a Hf-W age of 3.8±1.2 Ma after CAI formation.

Since the CH separates were separated according to their magnetic susceptibility, the linear correlation between $\varepsilon^{182}$W and $^{180}$Hf/$^{184}$W could also be a mixing line between W-rich and Hf-free metal and almost W-free and Hf-rich silicates. Such a mixing line may have no chronological significance if the metal and silicates were characterized by different initial $\varepsilon^{182}$W values or formed independently of each other at different times. However, several lines of evidence indicate that this is not the case. First, a mixing line would be evident by a linear correlation of the data in a diagram of $\varepsilon^{182}$W vs. 1/W (**Fig. 5b**). This is indeed the case, reflecting that the distinct Hf/W ratios of the Acfer 182 subsamples have different metal/silicate ratios, but the scatter of this correlation line (MSWD = 3.4) is much larger than that observed for the isochron (MSWD = 0.17). Much of this scatter results from the varying W concentrations of the different metal separates M-1 and M-2, but even excluding M-1 results in a regression line which still exhibits significantly more scatter (MSWD = 1.7) than the isochron. Second, bulk carbonaceous chondrites with a mean $^{180}$Hf/$^{184}$W ≈ 1.35 and $\varepsilon^{182}$W ≈ –1.9 plot on the Acfer 182 isochron. This composition represents the most appropriate starting composition for material in the CC reservoir, and so a mixing line between two components that formed at different times from a bulk carbonaceous chondrite reservoir would not pass through this composition. Finally, the combined model age of the CH metals of 3.8±1.5 Ma [2 s.d., $n$ = 2; **Table S3**] after CAI formation is in good agreement with the Hf-W isochron age for Acfer 182, indicating that CH silicates and metal formed about contemporaneously. Together, these observations indicate that the Hf-W data for metal and silicate separates from the CH chondrite Acfer 182 define a meaningful isochron, which provides a time of metal-silicate fractionation of 3.8±1.2 Ma after CAI formation.

Krot et al. (2017) reported Al-Mg data for individual porphyritic chondrules and plagioclase fragments from the paired CH chondrites Acfer 182 and 214, and found that the vast majority of the investigated samples show no resolved radiogenic $^{26}$Mg excesses, providing only a minimum age for chondrule formation of >3.5 Ma after CAI formation. The Hf-W age of 3.8±1.2 Ma after CAI formation is consistent with these Al-Mg results, and, unlike the Al-Mg data, also provides the first precise age for CH chondrites.

A key observation from the new Hf-W data is that CB metal ($\varepsilon^{182}W = -3.06 \pm 0.12$) plots precisely on the Hf-W isochron defined by metal and silicate fractions of the CH chondrite Acfer 182 (initial $\varepsilon^{182}W$ of the isochron is $-3.10 \pm 0.09$), indicating that metal and silicates in CB and CH chondrites formed at about the same time. This combined with the strong genetic link between CB and CH chondrites inferred from their nucleosynthetic isotope signatures suggests strongly that CB and CH chondrites formed together from the same building material. A weighted average Hf-W age from the isochron of Acfer 182 (CH) and the three model ages of Gujba (CB$_a$), Bencubbin (CB$_a$), and Isheyevo (CB$_b$/CH) of 3.8±0.8 Ma [95% CI, $n = 4$; **Fig. 7**] after CAI formation, thus, represents our best estimate for the formation time of CB/CH chondrites.

## 6   Implications for the origin of CB and CH chondrites

The isotopic and chronological results of this study have implications for understanding the formation of CB and CH chondrites, the origin of their precursor material, and their relation to CR chondrites. The key observations of this study may be summarized as follows: (*i*) CB and CH chondrites formed contemporaneously at ~3.8 Ma after CAI formation; CR chondrites also formed at about the same time (Nagashima et al., 2014; Schrader et al., 2017; Budde et al., 2018); (*ii*) CB and CH chondrites formed from the same metal and silicate components, which themselves have distinct isotopic compositions, and where CH chondrites incorporated a larger fraction of silicate material than the CB chondrites; (*iii*) CR, CB, and CH metal is isotopically indistinguishable, suggesting a common origin at least for their precursors, i.e., formation of these materials in the same region of the solar nebula. However, CR silicates are isotopically distinct from silicates in CB/CH chondrites, indicating that CR chondrites are not directly genetically related to CB and CH chondrites; (*iv*) metal and silicates in CH chondrites exhibit distinct Mo isotope anomalies, where the silicates are characterized by an *s*-process deficit compared to the metal.

*6.1   Origin of CB and CH chondrites*

The leading theory for the origin of CB and CH chondrites involves formation in an impact-induced vapor plume (e.g., Wasson and Kallemeyn, 1990; Krot et al., 2005; Olsen et al., 2013; Fedkin et al., 2015; Weyrauch et al., 2019; Koefoed et al., 2022). Within this framework, CB and CH chondrites may have formed from the same, compositionally heterogeneous vapor-melt plume, in which metal and silicate condensed contemporaneously but at different locations within the plume (Weyrauch et al., 2019, 2021). As such, CB chondrites may be samples from the innermost part of the impact plume, whereas CH chondrites derive from more distal regions of the plume, which was richer in silicate phases.

For the most part, the data of this study are consistent with this model. In particular, our finding that CB and CH chondrites formed contemporaneously and represent variable mixtures of the same metal and silicate components is consistent with the proposed formation of CB and CH chondrites in different regions of the same vapor plume. As noted above, CB and CH chondrites largely escaped modifications by parent body processes (e.g., Bischoff et al., 1993a, b; Krot et al., 2002), and so their Hf-W age provides the time of metal-silicate fractionation in the vapor plume during CB and CH chondrite formation. As condensation of the vapor plume is expected to have occurred rapidly after the impact (e.g., Fedkin et al., 2015), the Hf-W age of CB and CH chondrites most likely also provides the time of the impact, which according to our new data occurred at 3.8±0.8 Ma after CAI formation.

However, the distinct Mo isotopic compositions of metal and silicates in CH (and CB) chondrites are more difficult to understand within the framework of formation from a vapor plume. Due to fast diffusional exchange in the gas phase, metal and silicates which formed from a common vapor plume are expected to be in chemical and isotopic equilibrium. This is consistent with the Hf-W isotope systematics, which indicate isotopic equilibration between CH metal and silicates during their formation, but is more difficult to reconcile with the observation that CH metal is enriched in *s*-process Mo isotopes compared to CH silicates. These distinct Mo isotopic compositions are unlikely to reflect physical mixing of genetically unrelated materials, because the Hf-W systematics indicate contemporaneous formation of CH/CB metal and CH silicates, and because metal and silicates are thought to have formed together from a common vapor plume. As noted above (Section **4.1**), similar *s*-process Mo isotope heterogeneities have been observed between matrix and chondrules of CV chondrites (Budde et al., 2016a) and metal and silicates of CR chondrites (Budde et al., 2018). In particular for the CV chondrites, some studies have argued that these internal *s*-process variations are due to parent body processes (Alexander, 2019; Sanders and Scott, 2022), but such an origin is highly unlikely for the pristine CH chondrites. Instead, Budde et al. (2016a) and Budde et al. (2018) have argued that the *s*-process variations reflect the preferential incorporation of a presolar metallic *s*-process carrier into the matrix (CV chondrites) or metal (CR chondrites) during chondrule formation. However, both CV and CR chondrites are thought to have formed in a nebular setting, and so it is unclear as to whether a similar process can have also occurred during formation in an impact. In particular, unlike primitive dust in the solar nebula, a vapor plume is not *a priori* expected to contain presolar material, but the presence of such material is necessary if the *s*-process Mo isotope variations were produced during formation of metal and silicates in the vapor plume. Of note, CH chondrites and to a lesser extent CB chondrites contain relict materials (e.g., CAIs, dark inclusions) that are thought to have formed in the solar nebula prior to formation of the CB/CH chondrites themselves, indicating that not all material in the vapor plume has been processed and homogenized. Thus, one possibility is that either some presolar material survived homogenization during the impact and was then preferentially incorporated into the forming metal or that the more distal areas of the vapor plume incorporated some primitive dust that surrounded the vapor plume. The latter would be consistent with the idea that CH chondrites formed in more distal areas of the plume. Clearly, explaining the *s*-process variations among metal and silicates in CH chondrites within the vapor plume model is not straightforward and requires further investigation, but these variations do not seem to be inconsistent with an origin of the CH and CB chondrites in an impact-induced vapor plume.

*6.2  Precursor material of CB and CH chondrites*

All CB and CH chondrite samples of this study, even at the scale of individual components, plot along the CC-line in Mo isotope space (**Fig. 2**), consistent with a CC heritage of these chondrites. As such, these data rule out that the CH and CB chondrites incorporated significant amounts of material originally formed in the NC reservoir, i.e., inner Solar System. For instance, Fedkin et al. (2015) argued that an ordinary chondrite-like parent body could have been one of the two colliding planetesimals that have been involved in the formation of a vapor-melt plume from which the CB and CH metal and silicate later condensed. However, since ordinary chondrites belong to the NC group of meteorites, in this model CB and CH samples should exhibit Mo isotopic compositions intermediate between the NC and CC lines, which is not observed. As such, the new Mo isotope data indicate that the precursor material of the CB and CH chondrites entirely derives from the CC reservoir, meaning that the two colliding bodies were CC bodies. This does not necessarily mean that the collision between the two bodies also occurred in the CC reservoir, which is thought to have been located beyond the orbit of Jupiter.

In fact, the collision might be associated with the predicted inward scattering of CC bodies during the growth and/or migration of Jupiter (e.g., Johnson et al., 2016; Raymond and Izidoro, 2017).

The general bulk chemical, mineralogical, and O and N isotopic similarities between CB, CH and CR chondrites have led to the proposal that at least one of the two colliding bodies involved in the formation of CB and CH chondrites was chemically (and isotopically) similar to CR chondrites (e.g., Fedkin et al., 2015). Our finding of identical Mo isotopic signatures of metal from CB, CH and CR chondrites supports this view. However, we also find that the silicates from CB/CH and CR chondrites are isotopically distinct, indicating that a bulk CR-like planetary body cannot have been the sole precursor material of the CB and CH chondrites. Instead, one possibility is that only CR-like metal contributed to the material that formed the CB and CH chondrites. For instance, based on detailed equilibrium condensation calculations, Fedkin et al. (2015), Oulton et al. (2016), and Koefoed et al. (2022) proposed that the formation of CB and CH chondrites involved the collision between a metal-rich impactor and a silicate-rich, chemically differentiated target. This is consistent with the results of this study, with the important additional implication that only the metal-rich body was a CR chondrite-like body. By contrast, the silicate-rich object was likely isotopically distinct from CR chondrites, as is evident from the distinct Ti isotopic compositions of CR and CB/CH chondrites. It is important to recognize that the Mo isotopic composition of CR, CB, and CH metal (e.g., $\varepsilon^{94}$Mo ≈ 1.3) is not uncommon among CC meteorites and that many CC irons have a very similar Mo isotopic compositions (e.g., Kruijer et al., 2017; Worsham et al., 2019). As such, at least from a Mo isotopic standpoint, many of the CC irons would be suitable candidates for the metal-rich precursor body of the CB and CH chondrites. Nevertheless, CR chondrites are additionally linked to CB and CH chondrites by their anomalously high $^{15}$N enrichments, which among the CC irons is only found for the IIC irons. These, however, are characterized by a distinct Mo isotopic composition (e.g., $\varepsilon^{94}$Mo ≈ 2.2). As such, when the isotopic compositions of known CC meteorites are considered, CR chondrite-like metal itself seems to be the best candidate for the metal-rich precursor body of the CB and CH chondrites.

Finally, the Hf-W data suggest that CR and CB/CH chondrites seem to have undergone metal-silicate separation at about the same time. Although the uncertainties on their Hf-W ages (3.7±0.7 Ma for the CR chondrites, 3.8±0.8 Ma for the CB/CH chondrites) are sufficiently large to allow for significant age difference, the good agreement between these ages is quite striking and is reinforced by the observation that all these metals have indistinguishable $^{182}$W isotopic compositions. Thus, in addition to the close genetic link between CR and CB/CH chondrites, these samples also seem to be chronologically linked. However, such a chronological link is not expected, given that CR and CB/CH chondrites are thought to have formed by different processes and in different environments (solar nebula versus planetesimal impacts). Although striking, the chronological link between CR and CB/CH chondrites, therefore, seems to be coincidence.

## 7 Conclusions

The data of this study indicate that CB and CH chondrites formed contemporaneously at ~3.8 Ma after CAI formation, and predominantly consist of the same metal and silicate components. These components are isotopically distinct, where the silicates exhibit a larger *s*-process deficit than the metal, probably due to the preferential incorporation of an *s*-process carrier in the metal over the silicates. Metal in CB and CH has the same Mo isotopic composition as metal in CR chondrites, consistent with a common genetic origin. However, CR silicates have distinct Mo and Ti isotopic compositions compared to silicates in CB/CH chondrites.

Collectively, the new isotopic data are consistent with formation of CB and CH chondrites in different regions of the same impact-generated vapor-melt plume, where CH chondrites incorporated a larger fraction of silicate material and also a larger fraction of solar nebula material that had not been fully homogenized during impact. The combined Mo and Ti isotopic data indicate that the precursor bodies of the CB and CH chondrites were CC-type bodies, and that only the metal in these bodies is genetically linked to CR chondrites, whereas the silicate has a different origin. Together, this study demonstrates the importance of using nucleosynthetic isotope anomalies of different elements to disentangle the genetic relationship among the distinct components of meteorites, especially for samples where metal and silicates may have been mixed during the collision between objects with disparate compositions.

# 8   Acknowledgments

We gratefully acknowledge Addi Bischoff (University of Münster) for providing meteorite samples for this study as well as U. Heitmann for sample preparation. We also thank Ian Sanders and two anonymous reviewers for their constructive comments and Audrey Bouvier for efficient editorial handling. This work was funded by the Deutsche Forschungsgemeinschaft (DFG, German Research Foundation) – Project-ID 263649064 – TRR 170. The funding is gratefully acknowledged. This is TRR 170 publication no. 206.

# 9   Data Availability

Data are available through the TRR 170 Repository TRR170-DB at doi.org/10.35003/4NMC6P.

## 10  Appendix A. Supplementary Material

Supplementary material related to this article includes Hf-W and Ti isotope data for the geological reference materials and a summary of ages reported for CB, CH and CR chondrites using different chronometers.

nebular and regolithic heritage. Earth Planet. Sci. Lett. 101, 148–161.

Weisberg, M.K., Prinz, M., Clayton, R.N., Mayeda, T.K., Grady, M.M., Pillinger, C.T., 1995. The CR chondrite clan. Proc NIPR Symp Antarct Meteor. 8, 11–32.

Weisberg, M.K., Prinz, M., Clayton, R.N., Mayeda, T.K., Sugiura, N., Zashu, S., Ebihara, M., 2001. A new metal-rich chondrite grouplet. Meteorit. Planet. Sci. 36, 401–418.

Weyrauch, M., Zipfel, J., Weyer, S., 2019. Origin of metal from CB chondrites in an impact plume – A combined study of Fe and Ni isotope composition and trace element abundances. Geochim. Cosmochim. Acta 246, 123–137.

Weyrauch, M., Zipfel, J., Weyer, S., 2021. The relationship of CH, CB, and CR chondrites: Constraints from trace elements and Fe-Ni isotope systematics in metal. Geochim. Cosmochim. Acta 308, 291–309.

Worsham, E.A., Burkhardt, C., Budde, G., Fischer-Gödde, M., Kruijer, T.S., Kleine, T., 2019. Distinct evolution of the carbonaceous and non-carbonaceous reservoirs: Insights from Ru, Mo, and W isotopes. Earth Planet. Sci. Lett. 521, 103–112.

Yamashita, K., Maruyama, S., Yamakawa, A., Nakamura, E., 2010. $^{53}$Mn-$^{53}$Cr Chronometry of CB chondrite: Evidence for uniform distribution of $^{53}$Mn in the early Solar System. Astrophys. J. 723, 20–24.

Yin, Q., Jacobsen, S.B., Yamashita, K., Blichert-Toft, J., Télouk, P., Albarède, F., 2002. A short timescale for terrestrial planet formation from Hf–W chronometry of meteorites. Nature 418, 949–952.

Zhang, J., Dauphas, N., Davis, A.M., Pourmand, A., 2011. A new method for MC-ICPMS measurement of titanium isotopic composition: Identification of correlated isotope anomalies in meteorites. J. Anal. At. Spectrom. 26, 2197.

Zhang, J., Dauphas, N., Davis, A.M., Leya, I., Fedkin, A., 2012. The proto-Earth as a significant source of lunar material. Nat. Geosci. 5, 251–255.

Zhu, K., Moynier, F., Schiller, M., Alexander, C.M.O., Davidson, J., Schrader, D.L., van Kooten, E., Bizzarro, M., 2021. Chromium isotopic insights into the origin of chondrite parent bodies and the early terrestrial volatile depletion. Geochim. Cosmochim. Acta 301, 158–186.
38

**Table 1:** Molybdenum, Ti, and Cr isotope data for CB, CH and CR chondrite samples.

| Sample | $Mo^b$ (μg/g) | Mo/W | N (Mo-IC) | $\varepsilon^{92}Mo$ (± 2σ) | $\varepsilon^{94}Mo$ (± 2σ) | $\varepsilon^{95}Mo$ (± 2σ) | $\varepsilon^{97}Mo$ (± 2σ) | $\varepsilon^{100}Mo$ (± 2σ) | $Ti^b$ (μg/g) | N (Ti-IC) | $\varepsilon^{46}Ti$ (± 2σ) | $\varepsilon^{48}Ti$ (± 2σ) | $\varepsilon^{50}Ti$ (± 2σ) |
|---|---|---|---|---|---|---|---|---|---|---|---|---|---|
| *CH (Acfer 182 – CH3)* | | | | | | | | | | | | | |
| Bulk[a,c] | 1.60 | 9.1 | 6 | 2.32 ± 0.11 | 1.79 ± 0.10 | 1.29 ± 0.04 | 0.62 ± 0.09 | 0.66 ± 0.11 | 320 | 10 | 0.37 ± 0.08 | –0.08 ± 0.09 | 2.08 ± 0.05 |
| M-1 | 5.62 | 11.3 | 7 | 1.48 ± 0.08 | 1.21 ± 0.05 | 0.97 ± 0.03 | 0.49 ± 0.04 | 0.52 ± 0.12 | – | – | – | – | – |
| M-2 | 4.87 | 14.0 | 7 | 1.62 ± 0.08 | 1.29 ± 0.04 | 1.00 ± 0.08 | 0.52 ± 0.06 | 0.53 ± 0.09 | – | – | – | – | – |
| Wt. av. CH metal (*n* = 2) | 5.25 | 12.7 | – | 1.55 ± 0.14 | 1.26 ± 0.08 | 0.97 ± 0.03 | 0.50 ± 0.03 | 0.53 ± 0.07 | – | – | – | – | – |
| MS-1 | 3.12 | 13.4 | 7 | 1.87 ± 0.21 | 1.46 ± 0.06 | 1.12 ± 0.08 | 0.56 ± 0.04 | 0.58 ± 0.07 | 264 | 15 | 0.36 ± 0.08 | –0.12 ± 0.04 | 1.91 ± 0.10 |
| MS-2 | 2.54 | 12.8 | 7 | 2.04 ± 0.20 | 1.58 ± 0.06 | 1.20 ± 0.07 | 0.61 ± 0.04 | 0.64 ± 0.08 | 360 | 15 | 0.29 ± 0.07 | –0.11 ± 0.01 | 1.96 ± 0.07 |
| MS-3 | 1.21 | 9.4 | 5 | 2.65 ± 0.27 | 2.00 ± 0.11 | 1.45 ± 0.10 | 0.69 ± 0.05 | 0.79 ± 0.18 | 487 | 15 | 0.36 ± 0.05 | –0.10 ± 0.04 | 1.91 ± 0.09 |
| MS-4 | 0.96 | 8.8 | 5 | 3.05 ± 0.17 | 2.39 ± 0.09 | 1.63 ± 0.11 | 0.86 ± 0.05 | 1.00 ± 0.11 | 547 | 15 | 0.39 ± 0.08 | –0.07 ± 0.03 | 1.94 ± 0.09 |
| MS-5 | 0.66 | 7.3 | 4 | 3.26 ± 0.26 | 2.52 ± 0.13 | 1.79 ± 0.11 | 0.91 ± 0.11 | 0.95 ± 0.14 | 722 | 15 | 0.29 ± 0.09 | –0.04 ± 0.04 | 2.06 ± 0.08 |
| F-1 | 1.32 | 9.6 | – | – | – | – | – | – | 507 | 15 | 0.41 ± 0.07 | –0.11 ± 0.04 | 1.94 ± 0.06 |
| *CH/CB$_b$ (Isheyevo)* | | | | | | | | | | | | | |
| Metal | 4.41 | 14.3 | 8 | 1.64 ± 0.13 | 1.30 ± 0.08 | 1.00 ± 0.07 | 0.53 ± 0.04 | 0.50 ± 0.06 | – | – | – | – | – |
| *CB chondrites* | | | | | | | | | | | | | |
| HaH 237 (CB$_b$) bulk[c] | 3.23 | 10.9 | 9 | 1.53 ± 0.08 | 1.26 ± 0.04 | 0.99 ± 0.04 | 0.51 ± 0.04 | 0.45 ± 0.04 | 105 | 15 | 0.37 ± 0.09 | –0.17 ± 0.03 | 1.91 ± 0.08 |
| Gujba (CB$_a$) metal | 5.03 | 7.7 | 7 | 1.78 ± 0.17 | 1.29 ± 0.11 | 1.01 ± 0.06 | 0.53 ± 0.05 | 0.52 ± 0.10 | – | – | – | – | – |
| Gujba (CB$_a$) metal[d] | 4.47 | – | 7 | 1.74 ± 0.25 | 1.30 ± 0.26 | 0.95 ± 0.10 | 0.47 ± 0.09 | 0.59 ± 0.23 | – | – | – | – | – |
| Bencubbin (CB$_a$) metal | 5.12 | 9.4 | 6 | 1.79 ± 0.19 | 1.34 ± 0.09 | 1.01 ± 0.08 | 0.55 ± 0.06 | 0.56 ± 0.07 | – | – | – | – | – |
| Wt. av. CB metal (*n* = 3) | 4.87 | 8.6 | – | 1.78 ± 0.11 | 1.32 ± 0.07 | 1.00 ± 0.04 | 0.53 ± 0.04 | 0.55 ± 0.06 | – | – | – | – | – |
| *CR chondrites* | | | | | | | | | | | | | |
| Acfer 139 (CR2) bulk | 1.18 | 7.5 | 5 | 3.60 ± 0.11 | 2.70 ± 0.07 | 1.92 ± 0.11 | 1.00 ± 0.02 | 1.19 ± 0.19 | 637 | 11 | 0.44 ± 0.09 | –0.12 ± 0.05 | 2.56 ± 0.05 |
| NWA 801 (CR2.8) bulk[d,f] | 1.21 | 8.5 | 3 | 3.58 ± 0.64 | 2.82 ± 0.38 | 1.73 ± 0.32 | 0.87 ± 0.36 | 1.03 ± 0.11 | – | 15 | 0.47 ± 0.05 | –0.05 ± 0.12 | 2.35 ± 0.04 |
| GRA 06100 (CR2.5) bulk[e,g] | 1.44 | 8.4 | 5 | 4.14 ± 0.12 | 3.11 ± 0.15 | 2.26 ± 0.04 | 1.18 ± 0.04 | 1.40 ± 0.17 | 715 | 10 | 0.60 ± 0.10 | 0.04 ± 0.04 | 3.37 ± 0.09 |
| Wt. av. CR metal[e] (*n* = 5) | 4.71 | 8.5 | – | 1.81 ± 0.07 | 1.37 ± 0.05 | 1.04 ± 0.04 | 0.54 ± 0.03 | 0.58 ± 0.05 | – | – | – | – | – |

Mo, Ti, Cr isotope ratios are internally normalized to $^{98}Mo/^{96}Mo$ = 1.453173, $^{49}Ti/^{47}Ti$ = 0.749766, and $^{50}Cr/^{52}Cr$ = 0.051859, respectively. For samples with N>3, the uncertainties represent 95% confidence intervals (95% CI). N: number of analyses. M: metal, MS: magnetic separate ('mixed fractions'), F: fines.
[a] Bulk Acfer 182 has additionally been measured for its Cr isotopic composition (N = 9; [Cr]$^b$ = 2906 μg/g): $\varepsilon^{53}Cr$ = 0.25 ± 0.08 and $\varepsilon^{54}Cr$ = 1.41 ± 0.13.
[b] Mo, Ti, and Cr concentrations as determined by quadrupole ICP-MS, which have an uncertainty of ~5%.
[c] Mo concentration and Mo isotope data for bulk Acfer 182 and HaH 237 were reported in Budde et al. (2019).
[d] Mo concentration and Mo isotope data for Gujba metal and bulk NWA 801 are from Burkhardt et al. (2011).
[e] Mo concentration and Mo isotope data for bulk GRA 06100 and average CR metal are from Budde et al. (2018) and include five different digestions of metal separates.
[f] Ti isotope data for bulk NWA 801 are from Zhang et al. (2012).
[g] GRA 06100 seems chemically and isotopically disturbed (Budde et al., 2018; and references therein) and, thus, was not included into any calculations of average CR chondrite values (e.g., Ti isotopic composition; Section **3.2.**

**Table 2:** Hf-W isotope data for CB, CH and CR chondrite samples.

| Sample | Grain size (μm) | Weight (mg) | Hf (ng/g) | W (ng/g) | $^{180}$Hf/$^{184}$W (± 2σ) | N (W-IC) | ε$^{182}$W$_{meas.}$ (± 2σ) | ε$^{183}$W (± 2σ) | ε$^{182}$W$_i$ (± 2σ) | ε$^{182}$W$_{nuc.\ corr.}$ (± 2σ) | ε$^{182}$W$_{meas.}$ (± 2σ) | ε$^{184}$W (± 2σ) | ε$^{182}$W$_i$ (± 2σ) | ε$^{182}$W$_{nuc.\ corr.}$ (± 2σ) |
|---|---|---|---|---|---|---|---|---|---|---|---|---|---|---|
| | | | | | | | normalized to $^{186}$W/$^{184}$W = 0.92767 ('6/4') | | | | normalized to $^{186}$W/$^{183}$W = 1.98590 ('6/3') | | | |
| *CH (Acfer 182 – CH3)* | | | | | | | | | | | | | | |
| Bulk | – | 553 | 129 | 175 | 0.868 ± 0.004 | 2 | −2.23 ± 0.12 | 0.12 ± 0.16 | −2.99 ± 0.13 | −2.38 ± 0.23 | −2.39 ± 0.18 | −0.08 ± 0.11 | −3.15 ± 0.19 | −2.38 ± 0.19 |
| M-1 | >125 | 174 | 4 | 499 | 0.010 ± 0.002 | 2 | −2.92 ± 0.12 | 0.10 ± 0.16 | −2.93 ± 0.12 | −3.04 ± 0.23 | −3.04 ± 0.18 | −0.07 ± 0.11 | −3.05 ± 0.18 | −3.03 ± 0.19 |
| M-2 | 40-125 | 307 | 8 | 348 | 0.026 ± 0.002 | 3 | −2.93 ± 0.12 | 0.12 ± 0.16 | −2.95 ± 0.12 | −3.08 ± 0.23 | −3.08 ± 0.18 | −0.08 ± 0.11 | −3.11 ± 0.18 | −3.08 ± 0.19 |
| Wt. av. CH metal (*n* = 2) | – | – | 6 | 424 | 0.018 ± 0.016 | – | −2.93 ± 0.09 | 0.11 ± 0.11 | −2.94 ± 0.09 | −3.06 ± 0.16 | −3.06 ± 0.13 | −0.08 ± 0.08 | −3.08 ± 0.13 | −3.06 ± 0.13 |
| MS-1 | 40-125 | 411 | 72 | 233 | 0.365 ± 0.001 | 2 | −2.58 ± 0.12 | 0.16 ± 0.16 | −2.90 ± 0.12 | −2.77 ± 0.23 | −2.80 ± 0.18 | −0.10 ± 0.11 | −3.12 ± 0.19 | −2.79 ± 0.19 |
| MS-2 | 40-125 | 455 | 97 | 199 | 0.572 ± 0.002 | 2 | −2.35 ± 0.12 | 0.18 ± 0.16 | −2.86 ± 0.12 | −2.58 ± 0.23 | −2.59 ± 0.18 | −0.12 ± 0.11 | −3.09 ± 0.19 | −2.58 ± 0.19 |
| MS-3 | 40-125 | 501 | 154 | 129 | 1.413 ± 0.004 | 2 | −1.65 ± 0.12 | 0.12 ± 0.16 | −2.89 ± 0.15 | −1.80 ± 0.23 | −1.81 ± 0.18 | −0.08 ± 0.11 | −3.05 ± 0.21 | −1.80 ± 0.19 |
| MS-4 | 40-125 | 540 | 172 | 109 | 1.865 ± 0.005 | 1 | −1.20 ± 0.12 | 0.22 ± 0.16 | −2.84 ± 0.17 | −1.47 ± 0.23 | −1.49 ± 0.18 | −0.14 ± 0.11 | −3.13 ± 0.23 | −1.47 ± 0.19 |
| MS-5 | 40-125 | 542 | 216 | 91 | 2.806 ± 0.011 | 1 | −0.51 ± 0.12 | 0.10 ± 0.16 | −2.97 ± 0.23 | −0.63 ± 0.23 | −0.62 ± 0.18 | −0.07 ± 0.11 | −3.08 ± 0.27 | −0.61 ± 0.19 |
| F-1 | <40 | 557 | 149 | 138 | 1.280 ± 0.004 | 2 | −1.71 ± 0.12 | 0.22 ± 0.16 | −2.83 ± 0.15 | −1.98 ± 0.23 | −2.00 ± 0.18 | −0.15 ± 0.11 | −3.13 ± 0.21 | −1.99 ± 0.19 |
| *CH/CB$_b$ (Isheyevo)* | | | | | | | | | | | | | | |
| Metal | – | 489 | 17 | 308 | 0.065 ± 0.003 | 5 | −2.81 ± 0.07 | 0.17 ± 0.14 | −2.87 ± 0.07 | −3.02 ± 0.19 | −3.04 ± 0.14 | −0.11 ± 0.10 | −3.09 ± 0.14 | −3.02 ± 0.14 |
| *CB chondrites* | | | | | | | | | | | | | | |
| HaH 237 (CB$_b$) bulk | – | 561 | 54 | 297 | 0.214 ± 0.001 | 4 | −2.71 ± 0.05 | 0.10 ± 0.14 | −2.89 ± 0.05 | −2.83 ± 0.19 | −2.84 ± 0.16 | −0.07 ± 0.10 | −3.03 ± 0.16 | −2.83 ± 0.16 |
| Gujba (CB$_a$) metal | – | 403 | 3 | 651 | 0.006 ± 0.002 | 6 | −2.89 ± 0.03 | 0.11 ± 0.14 | −2.90 ± 0.03 | −3.03 ± 0.17 | −3.04 ± 0.15 | −0.07 ± 0.09 | −3.05 ± 0.15 | −3.03 ± 0.15 |
| Bencubbin (CB$_a$) metal | – | 166 | 16 | 542 | 0.034 ± 0.002 | 2 | −2.93 ± 0.12 | 0.13 ± 0.16 | −2.96 ± 0.12 | −3.09 ± 0.23 | −3.10 ± 0.18 | −0.09 ± 0.11 | −3.13 ± 0.18 | −3.09 ± 0.19 |
| Wt. av. CB metal (*n* = 2) | – | – | 10 | 597 | 0.020 ± 0.028 | – | −2.89 ± 0.03 | 0.12 ± 0.11 | −2.90 ± 0.03 | −3.05 ± 0.14 | −3.06 ± 0.12 | −0.08 ± 0.07 | −3.08 ± 0.12 | −3.05 ± 0.12 |
| *CR chondrites* | | | | | | | | | | | | | | |
| Acfer 139 (CR2) bulk | – | 538 | 173 | 158 | 1.295 ± 0.004 | 2 | −1.19 ± 0.12 | 0.45 ± 0.16 | −2.34 ± 0.12 | −1.75 ± 0.23 | −1.79 ± 0.18 | −0.30 ± 0.11 | −2.94 ± 0.19 | −1.76 ± 0.19 |
| NWA 801 (CR2.8) bulk[a] | – | 556 | 184 | 142 | 1.533 ± 0.011 | 2 | −1.08 ± 0.14 | 0.49 ± 0.16 | −2.44 ± 0.14 | −1.69 ± 0.24 | −1.74 ± 0.22 | −0.32 ± 0.11 | −3.10 ± 0.22 | −1.70 ± 0.22 |
| NWA 1180 (CR2) bulk[a] | – | 498 | 147 | 129 | 1.339 ± 0.012 | 1 | −1.52 ± 0.14 | 0.31 ± 0.16 | −2.71 ± 0.14 | −1.91 ± 0.24 | −1.94 ± 0.22 | −0.20 ± 0.11 | −3.13 ± 0.22 | −1.92 ± 0.22 |
| GRA 06100 (CR2.5) bulk[a] | – | 555 | 194 | 172 | 1.331 ± 0.004 | 3 | −1.50 ± 0.14 | 0.35 ± 0.16 | −2.68 ± 0.14 | −1.93 ± 0.24 | −1.96 ± 0.22 | −0.23 ± 0.11 | −3.14 ± 0.22 | −1.94 ± 0.22 |
| Wt. av. CR metal[a] (*n* = 11) | – | – | 5 | 506 | 0.015 ± 0.005 | – | −2.98 ± 0.06 | 0.18 ± 0.05 | −2.99 ± 0.06 | −3.18 ± 0.06 | −3.18 ± 0.06 | −0.12 ± 0.03 | −3.17 ± 0.06 | −3.17 ± 0.06 |

W isotope ratios that involve $^{183}$W were corrected for a small analytical $^{183}$W effect after Kruijer et al. (2014). Given uncertainties are based on the external reproducibility (2 s.d.) obtained from repeated analyses of the terrestrial standards (**Table S1**) or the in-run error (2 s.e.), whichever is larger. For samples with N>3, the uncertainties represent 95% confidence intervals (95% CI). Note that the final uncertainties include all propagated uncertainties induced by the correction for the $^{183}$W effect as well as by the correction for $^{182}$Hf decay (ε$^{182}$W$_i$) or nucleosynthetic anomalies (ε$^{182}$W$_{nuc.\ corr.}$) as described by Budde et al. (2016b). The initial $^{182}$Hf/$^{180}$Hf from the CH isochron was used to correct the analyzed CB and CH samples for $^{182}$Hf decay, whereas the initial $^{182}$Hf/$^{180}$Hf from the combined CR isochron reported by Budde et al. (2018) was used to correct the CR samples. N: number of analyses. M: metal, MS: magnetic separate ('mixed fractions'), F: fines.
[a] Hf-W concentrations and isotope data for bulk NWA 801, NWA 1180, GRA 06100, and average CR metal are from Budde et al. (2018) and include eleven different digestions of metal separates.

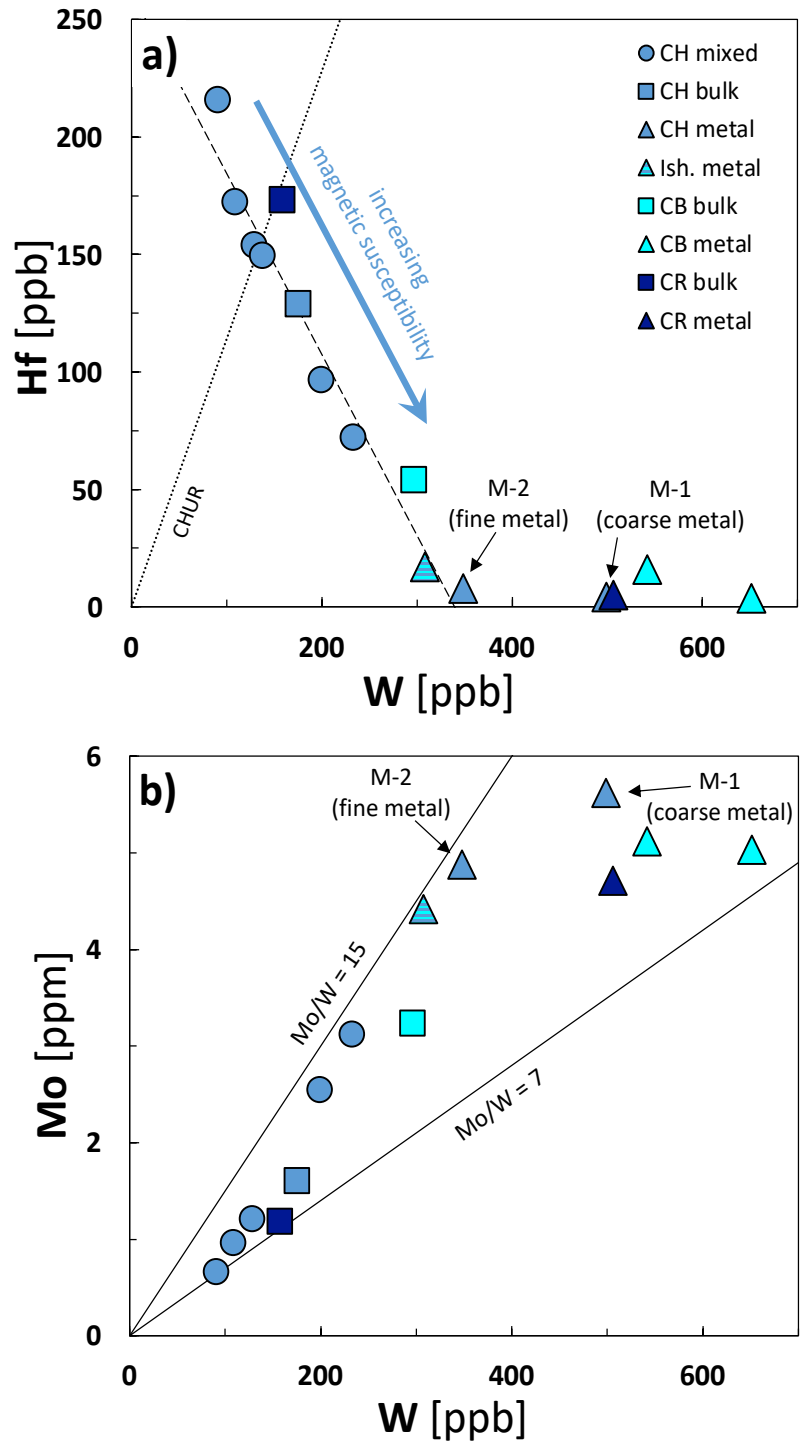

**Fig. 1:** (a) W vs. Hf and (b) W vs. Mo concentration diagram for the investigated CB, CH and CR chondrite samples. Note that Hf, Mo, and W contents of the CH mixed fractions (circles) vary significantly, which is attributable to the magnetic separation and the enrichment or depletion of W- and Mo-rich metal or Hf-rich silicate in the individual fractions. Interestingly, the bulk sample, the metal-silicate mixed fractions and the fine-grained metal separate M-2 of Acfer 182 (CH, medium blue symbols) plot on a single correlation line in Hf vs. W and Mo

vs. W space, respectively, which would be consistent with binary mixing between metal and silicate. Chondritic (CHUR) Hf-W data are adapted from Kleine et al. (2004). Mixed: metal-silicate separates MS-n.

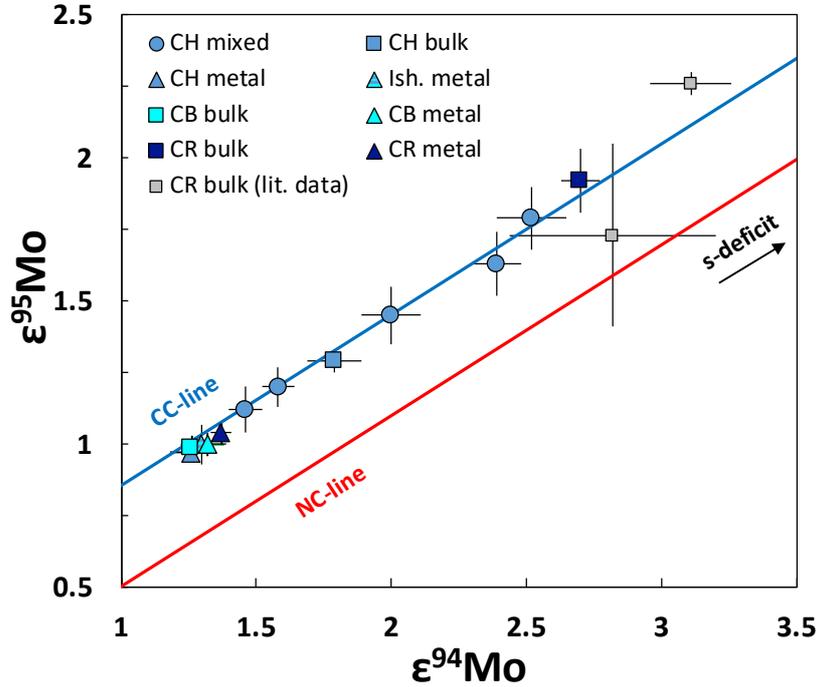

**Fig. 2:** Diagram of $\varepsilon^{94}$Mo vs. $\varepsilon^{95}$Mo for the analyzed CB, CH and CR chondrite samples. Bulk CB and CH chondrite data as well as slopes and intercepts of s-mixing lines as defined by carbonaceous (CC-line, blue) and non-carbonaceous meteorite samples (NC-line, red) are from Budde et al. (2019). Bulk CR data are from this study (Acfer 139, dark blue square) and from the literature (NWA 801 and GRA 06100; gray squares; Budde et al., 2018). Note that all samples from the present study plot along the CC-line, consistent with an outer Solar System origin of the metal-rich carbonaceous chondrites and, beyond that, excluding that these meteorites accreted significant amounts of material originally formed in the inner Solar System. Interestingly, mixed fractions and bulk samples of CB, CH and CR chondrites show systematically different Mo isotope compositions, however, their metals display identical $\varepsilon^{i}$Mo (e.g., $\varepsilon^{94}$Mo ≈ 1.3).

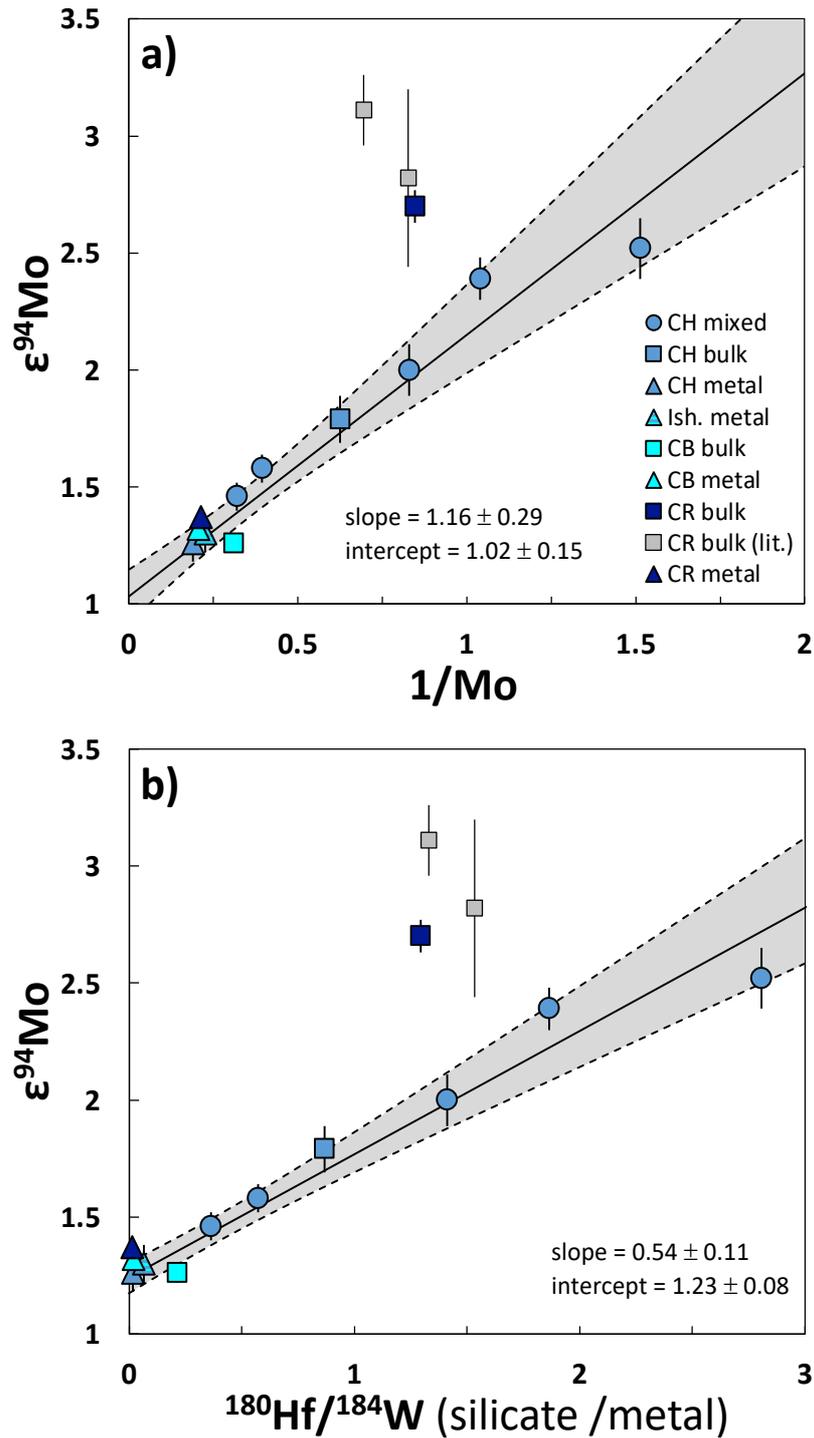

**Fig. 3:** Diagrams of (a) 1/Mo vs. $\varepsilon^{94}$Mo and (b) $^{180}$Hf/$^{184}$W vs. $\varepsilon^{94}$Mo for CB, CH and CR chondrite samples investigated in the present study, indicating that the $\varepsilon^{94}$Mo anomalies are correlated with silicate-to-metal ratio of the individual samples. Hence, nucleosynthetic Mo isotope anomalies scale with decreasing magnetic susceptibility. The solid black lines are linear regressions for the investigated CB and CH samples ($n = 10$, average CB and CH metal, respectively), calculated using *Isoplot*. Error envelopes of the regressions reflect 95% confidence intervals (95% CI). Bulk CR data (dark blue square from this study (Acfer 139); gray squares (NWA 801 and GRA 06100) from Burkhardt et al. (2011) and Budde et al. (2018), respectively) are shown for comparison and do not fall on the 1/Mo vs. $\varepsilon^{94}$Mo and $^{180}$Hf/$^{184}$W vs. $\varepsilon^{94}$Mo trends defined by the CB and CH samples.

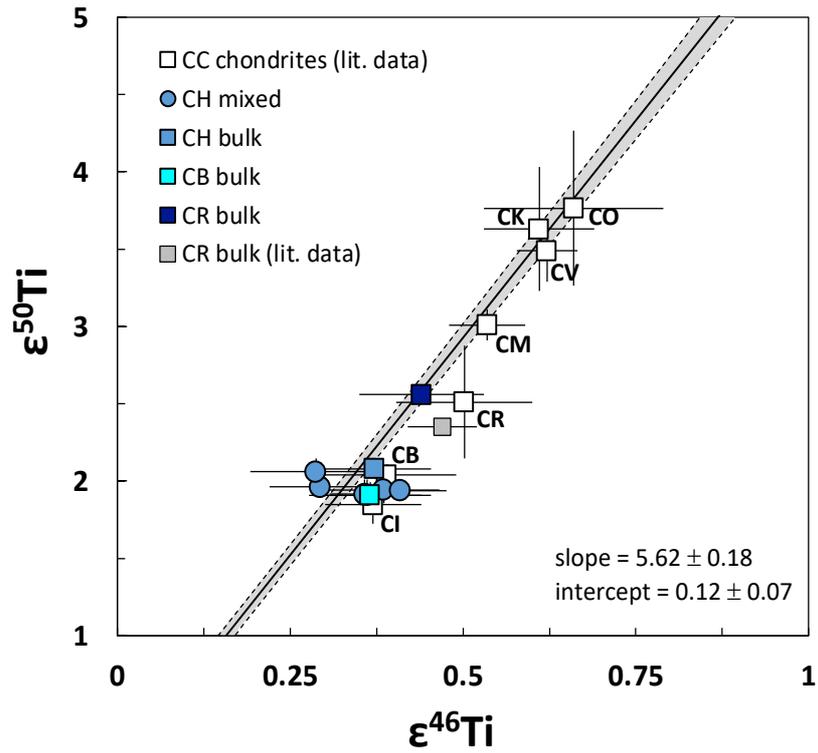

**Fig. 4:** Diagram of $\varepsilon^{46}$Ti vs. $\varepsilon^{50}$Ti for CB, CH and CR chondrite samples investigated in the present study. Bulk CR data for Acfer 139 (dark blue square) and NWA 801 (gray square) are from this study and from Zhang et al. (2012), respectively. Literature data for different carbonaceous chondrite groups are stated as composite points averaging multiple samples within each group and are shown for comparison (white squares). The weighted $\varepsilon^{50}$Ti–$\varepsilon^{46}$Ti regression (solid black line) and related error envelopes were calculated using *Isoplot*, and include bulk meteorite data available from previous works. All reported uncertainties reflect 95% confidence intervals (95% CI). Literature data were taken from the recent compilation of Burkhardt et al. (2019) and references therein.

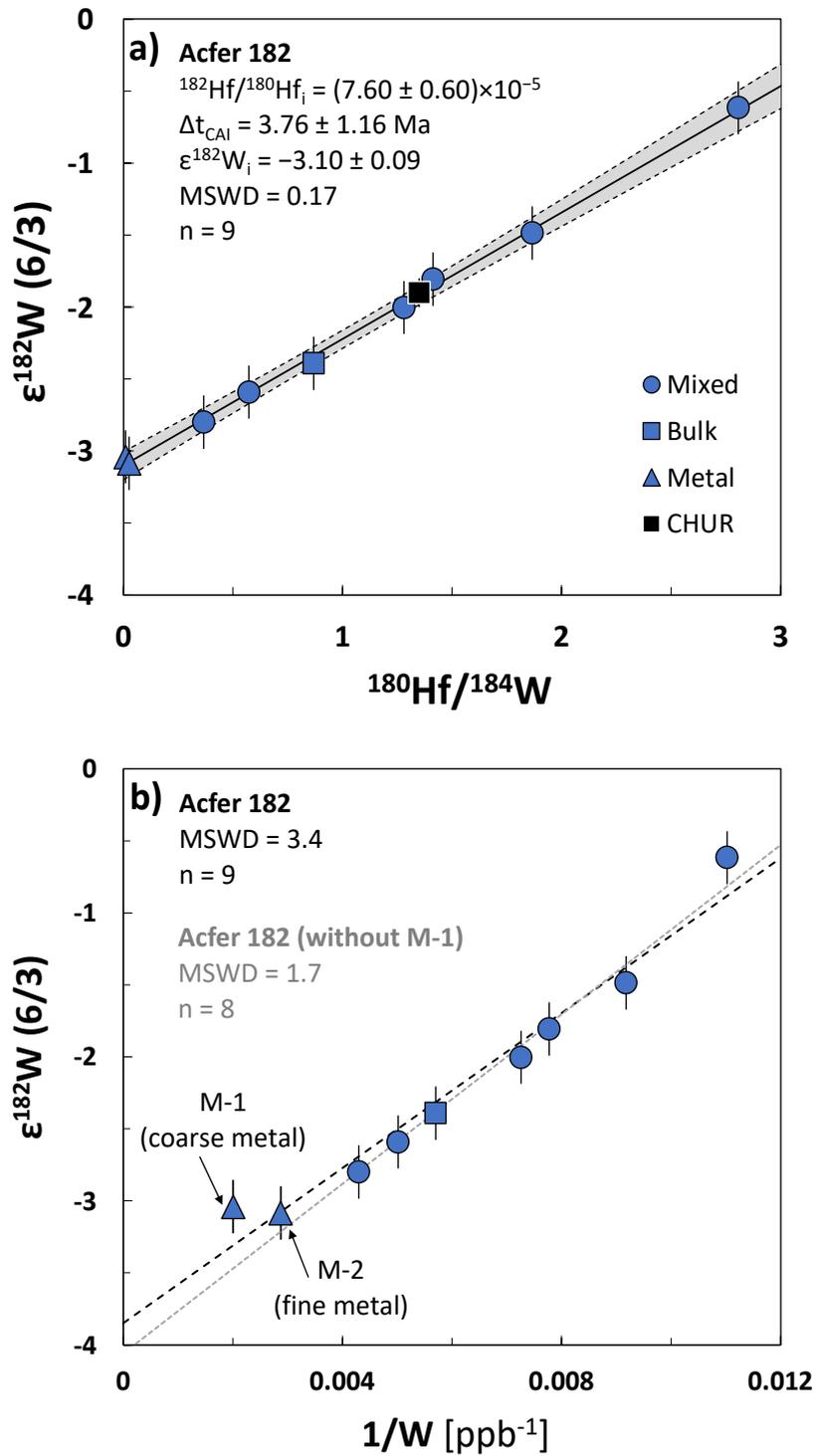

**Fig. 5:** (a) Internal Hf-W isochron diagram for Acfer 182 (CH). The $\varepsilon^{182}W$ values are normalized to $^{186}W/^{183}W$ ('6/3') and corrected for a small analytical $^{183}W$ effect (see section **2.2** for details). Note that bulk carbonaceous chondrites with a mean $^{180}Hf/^{184}W = 1.35$ and $\varepsilon^{182}W = -1.9 \pm 0.1$ (black square; CHUR; Kleine et al., 2002, 2004; Yin et al., 2002; Schoenberg et al., 2002) plot on the isochron (not included in the regression). (b) 1/W vs. $\varepsilon^{182}W$ (6/3) plot indicating that the different CH fractions probably reflect two-component mixing between a metal and a silicate endmember. All reported uncertainties are 95% confidence intervals (95% CI). $i$ = initial, $\Delta t_{CAI}$ = age relative to CAI formation, MSWD = mean square of weighted deviates, $n$ = number of samples included in the regression.

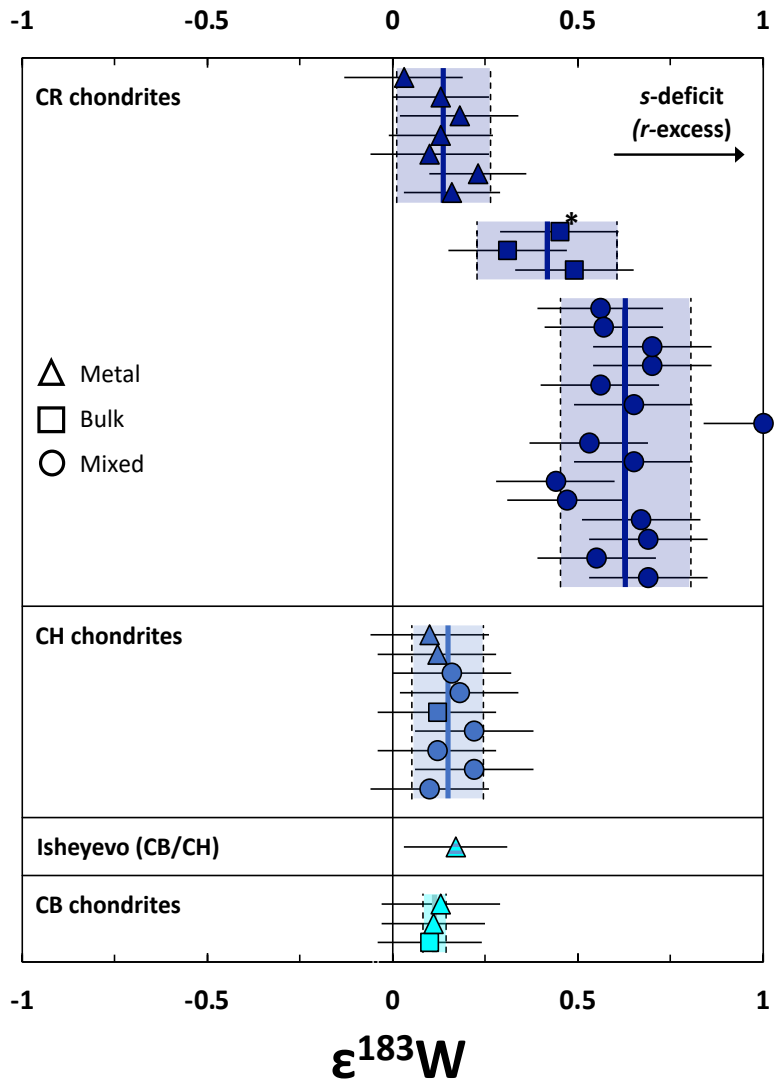

**Fig. 6:** Comparison of $\varepsilon^{183}W$ of bulk samples and individual fractions from CB, CH and CR chondrites. Thin vertical lines represent the mean $\varepsilon^{183}W$ of each chondrite group or displayed meteorite fraction (e.g., metal and silicates); transparent envelopes state the 2 s.d. of each sample type. All investigated CB and CH samples as well as CR metals exhibit small uniform excesses in $\varepsilon^{183}W$, which imply a deficit in s-process (or an excess in r-process) W isotopes. By contrast, bulk CR samples and CR mixed separates show larger and more variable $\varepsilon^{183}W$ excesses. CR data are from Budde et al. (2018) and references therein, except for bulk Acfer 139 (*, this study).

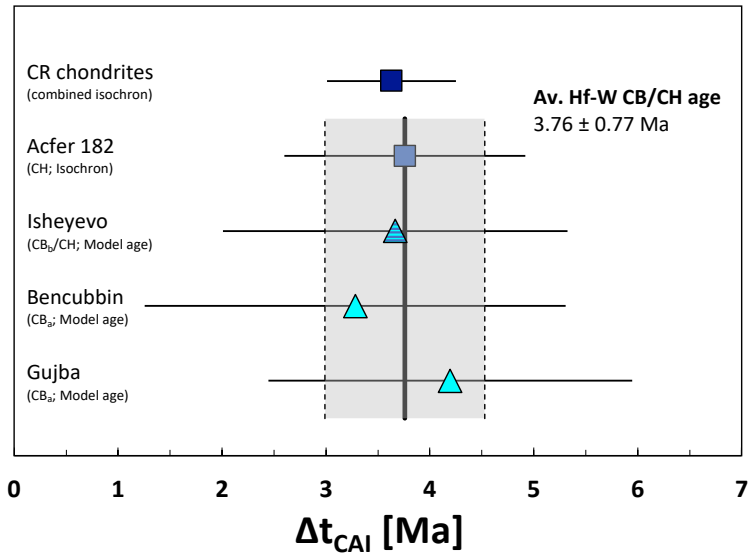

**Fig. 7:** Comparison of Hf-W ages obtained in this study for CB chondrites (model ages of Gujba and Bencubbin metal), Isheyevo (modal age of metal), and CH chondrites (isochron age of Acfer 182). Additionally, the Hf-W age reported for CR chondrites by Budde et al. (2018) is shown. The thin vertical line represents the average Hf-W age of CB and CH chondrites; the transparent envelope states the 2 s.d. of this age.

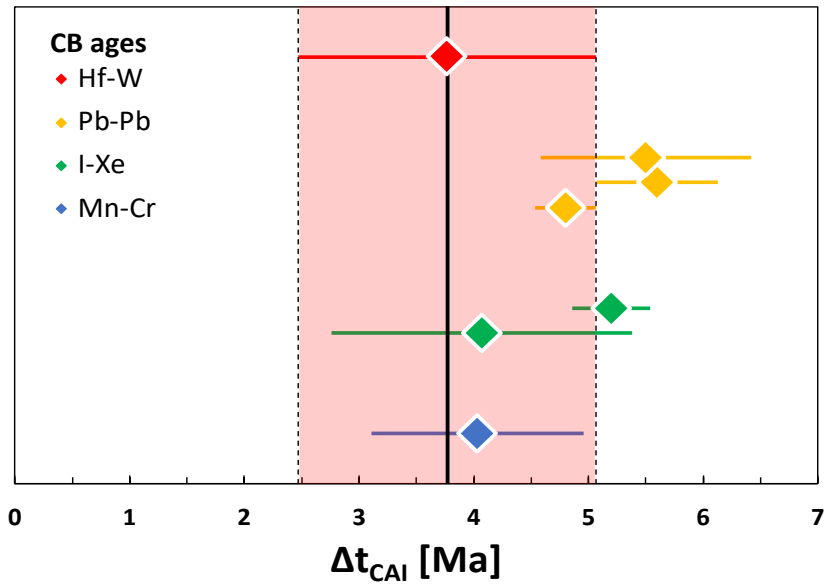

**Fig. 8:** Comparison of CB chondrite formation ages obtained by different chronometers. Hf-W ages provide the time of metal-silicate separation, Pb-Pb, Mn-Cr and I-Xe ages refer to the time of chondrule formation. Whereas the average Hf-W model age (this study), the Mn-Cr age (Yamashita et al., 2010), and the I-Xe age reported by Gilmour et al. (2009) are in excellent agreement with each other, the Pb-Pb ages (Krot et al., 2005; Bollard et al., 2015) and the I-Xe age reported by Pravdivtseva et al. (2017) seem to be slightly younger. See main text for discussion. Mn-Cr, I-Xe, and Pb-Pb ages were corrected for erroneously assumed $^{238}U/^{235}U = 137.88$ of their respective time anchors, using appropriate $^{238}U/^{235}U$ (see section **5.1** for details and references). Note that Al-Mg data for $CB_a$ and $CB_b$ chondrules reveal no radiogenic $^{26}Mg$ excesses (Gounelle et al., 2007; Olsen et al., 2013;

Nagashima et al., 2018), indicating they formed later than ~3.5 Ma after CAI formation, in line with the overall 'late' ages obtained for CB chondrites by the aforementioned chronometers.

Supplementary Material

# Age and genetic relationships among CB, CH and CR chondrites

Elias Wölfer, Gerrit Budde, and Thorsten Kleine

This file includes:
Supplementary Tables 1–3

**Table S1:** Hf-W isotope data for the geological reference materials.

**BHVO-2 (USGS), Hawaiian basalt**

| ID | Hf (ng/g) | W (ng/g) | $\varepsilon^{182}W_{meas.}$ (± 2 s.e.) | $\varepsilon^{183}W_{meas.}$ (± 2 s.e.) | $\varepsilon^{183}W_{corr.}$* (± 2σ) | $\varepsilon^{182}W_{meas.}$ (± 2 s.e.) | $\varepsilon^{182}W_{corr.}$* (± 2σ) | $\varepsilon^{184}W_{meas.}$ (± 2 s.e.) | $\varepsilon^{184}W_{corr.}$* (± 2σ) |
|---|---|---|---|---|---|---|---|---|---|
| | | | | normalized to $^{186}W/^{184}W$ = 0.92767 | | | normalized to $^{186}W/^{183}W$ = 1.98590 | | |
| BHV22.1 | 4580 | 221 | 0.00 ± 0.11 | -0.12 ± 0.10 | 0.00 ± 0.12 | 0.15 ± 0.10 | -0.01 ± 0.16 | 0.08 ± 0.07 | 0.00 ± 0.08 |
| BHV22.2 | – | – | 0.11 ± 0.10 | -0.02 ± 0.09 | 0.09 ± 0.11 | 0.14 ± 0.09 | 0.11 ± 0.15 | 0.01 ± 0.06 | -0.06 ± 0.07 |
| BHV22.3 | – | – | -0.05 ± 0.09 | -0.06 ± 0.08 | -0.05 ± 0.13 | 0.01 ± 0.08 | -0.07 ± 0.13 | 0.04 ± 0.05 | 0.03 ± 0.07 |
| BHV23.1 | – | – | -0.11 ± 0.10 | -0.23 ± 0.09 | -0.08 ± 0.11 | 0.20 ± 0.09 | -0.11 ± 0.15 | 0.16 ± 0.06 | 0.05 ± 0.07 |
| BHV23.2 | – | – | -0.07 ± 0.11 | -0.15 ± 0.09 | -0.05 ± 0.12 | 0.13 ± 0.09 | -0.07 ± 0.16 | 0.10 ± 0.06 | 0.04 ± 0.08 |
| BHV23.3 | – | – | 0.08 ± 0.09 | -0.06 ± 0.08 | 0.06 ± 0.10 | 0.16 ± 0.09 | 0.08 ± 0.14 | 0.04 ± 0.05 | -0.04 ± 0.07 |
| BHV24.1 | – | – | -0.01 ± 0.10 | -0.11 ± 0.10 | 0.00 ± 0.12 | 0.14 ± 0.09 | 0.00 ± 0.16 | 0.07 ± 0.06 | 0.00 ± 0.08 |
| BHV24.2 | – | – | 0.02 ± 0.10 | -0.12 ± 0.09 | 0.01 ± 0.11 | 0.17 ± 0.09 | 0.02 ± 0.15 | 0.08 ± 0.06 | -0.01 ± 0.07 |
| BHV24.3 | – | – | 0.00 ± 0.09 | -0.19 ± 0.07 | 0.01 ± 0.09 | 0.27 ± 0.08 | 0.02 ± 0.12 | 0.13 ± 0.05 | -0.01 ± 0.06 |
| BHV24.4 | – | – | 0.01 ± 0.10 | -0.22 ± 0.11 | -0.01 ± 0.13 | 0.28 ± 0.11 | -0.01 ± 0.17 | 0.14 ± 0.06 | 0.01 ± 0.08 |
| BHV25.1 | 4553 | 218 | 0.08 ± 0.10 | -0.04 ± 0.08 | 0.05 ± 0.10 | 0.12 ± 0.08 | 0.06 ± 0.13 | 0.03 ± 0.05 | -0.03 ± 0.07 |
| BHV25.2 | – | – | 0.01 ± 0.10 | -0.19 ± 0.08 | 0.01 ± 0.10 | 0.26 ± 0.08 | 0.01 ± 0.13 | 0.13 ± 0.05 | 0.00 ± 0.07 |
| BHV25.3 | – | – | -0.02 ± 0.09 | -0.10 ± 0.09 | -0.03 ± 0.11 | 0.08 ± 0.08 | -0.04 ± 0.14 | 0.06 ± 0.06 | 0.02 ± 0.07 |
| BHV26.1 | 4588 | 218 | 0.10 ± 0.09 | -0.07 ± 0.08 | 0.06 ± 0.10 | 0.17 ± 0.08 | 0.08 ± 0.13 | 0.05 ± 0.05 | -0.04 ± 0.06 |
| BHV26.2 | – | – | -0.03 ± 0.08 | -0.07 ± 0.08 | -0.02 ± 0.10 | 0.07 ± 0.08 | -0.02 ± 0.13 | 0.05 ± 0.05 | 0.01 ± 0.07 |
| BHV26.3 | – | – | -0.04 ± 0.09 | -0.09 ± 0.08 | 0.00 ± 0.09 | 0.13 ± 0.07 | 0.00 ± 0.13 | 0.06 ± 0.05 | 0.00 ± 0.06 |
| BHV26.4 | – | – | 0.00 ± 0.09 | -0.02 ± 0.08 | 0.01 ± 0.10 | 0.04 ± 0.08 | 0.02 ± 0.14 | 0.01 ± 0.05 | -0.01 ± 0.07 |
| BHV26.5 | – | – | -0.08 ± 0.09 | -0.16 ± 0.08 | -0.07 ± 0.10 | 0.13 ± 0.08 | -0.09 ± 0.14 | 0.11 ± 0.05 | 0.04 ± 0.07 |
| BHV26.6 | – | – | 0.07 ± 0.09 | 0.03 ± 0.08 | 0.05 ± 0.11 | 0.03 ± 0.09 | 0.07 ± 0.14 | -0.02 ± 0.06 | -0.04 ± 0.07 |
| BHV26.7 | – | – | -0.04 ± 0.10 | -0.20 ± 0.08 | -0.05 ± 0.11 | 0.21 ± 0.10 | -0.06 ± 0.14 | 0.13 ± 0.05 | 0.03 ± 0.07 |
| BHV27.1 | 4420 | 211 | 0.02 ± 0.10 | -0.11 ± 0.09 | 0.01 ± 0.11 | 0.15 ± 0.09 | 0.01 ± 0.15 | 0.07 ± 0.06 | 0.00 ± 0.08 |
| BHV27.2 | – | – | 0.04 ± 0.09 | -0.14 ± 0.08 | 0.03 ± 0.11 | 0.23 ± 0.09 | 0.04 ± 0.14 | 0.09 ± 0.05 | -0.02 ± 0.07 |
| BHV27.3 | – | – | -0.04 ± 0.10 | -0.20 ± 0.09 | -0.03 ± 0.12 | 0.23 ± 0.10 | -0.03 ± 0.16 | 0.13 ± 0.06 | 0.02 ± 0.08 |
| BHV28.1 | 4488 | 214 | 0.03 ± 0.11 | -0.07 ± 0.10 | 0.02 ± 0.12 | 0.12 ± 0.09 | 0.03 ± 0.16 | 0.04 ± 0.07 | -0.02 ± 0.08 |
| BHV28.2 | – | – | 0.04 ± 0.09 | -0.14 ± 0.08 | 0.01 ± 0.10 | 0.20 ± 0.09 | 0.02 ± 0.13 | 0.09 ± 0.05 | -0.01 ± 0.07 |
| BHV28.3 | – | – | -0.07 ± 0.10 | -0.24 ± 0.09 | -0.05 ± 0.11 | 0.25 ± 0.09 | -0.07 ± 0.14 | 0.16 ± 0.06 | 0.04 ± 0.07 |
| BHV29.1 | 4509 | 215 | -0.07 ± 0.10 | -0.17 ± 0.09 | -0.05 ± 0.10 | 0.15 ± 0.09 | -0.07 ± 0.14 | 0.11 ± 0.06 | 0.03 ± 0.07 |
| BHV29.2 | – | – | 0.01 ± 0.09 | -0.08 ± 0.09 | 0.01 ± 0.10 | 0.11 ± 0.08 | 0.01 ± 0.14 | 0.05 ± 0.06 | 0.00 ± 0.07 |
| BHV29.3 | – | – | -0.05 ± 0.09 | -0.07 ± 0.08 | -0.03 ± 0.10 | 0.06 ± 0.08 | -0.04 ± 0.13 | 0.05 ± 0.05 | 0.02 ± 0.07 |
| BHV30.1 | 4555 | 217 | -0.09 ± 0.09 | -0.13 ± 0.08 | -0.06 ± 0.10 | 0.09 ± 0.08 | -0.08 ± 0.13 | 0.09 ± 0.05 | 0.04 ± 0.06 |
| BHV30.2 | – | – | 0.00 ± 0.11 | -0.09 ± 0.09 | 0.00 ± 0.10 | 0.12 ± 0.09 | 0.00 ± 0.14 | 0.06 ± 0.06 | 0.00 ± 0.07 |
| BHV30.3 | – | – | 0.05 ± 0.10 | -0.10 ± 0.08 | 0.06 ± 0.10 | 0.21 ± 0.08 | 0.08 ± 0.13 | 0.07 ± 0.05 | -0.04 ± 0.07 |
| N | 7 | 7 | 32 | 32 | 32 | 32 | 32 | 32 | 32 |
| Mean | 4527 | 216 | 0.00 | -0.12 | 0.00 | 0.15 | 0.00 | 0.08 | 0.00 |
| 2 s.d. | 119 | 6.4 | 0.11 | 0.13 | 0.09 | 0.14 | 0.11 | 0.09 | 0.06 |
| 95% CI | 55 | 3.0 | 0.02 | 0.02 | 0.02 | 0.03 | 0.02 | 0.02 | 0.01 |

**SRM 129c (NIST), high-sulfur steel**

| ID | Hf (ng/g) | W (ng/g) | $\varepsilon^{182}W_{meas.}$ (± 2 s.e.) | $\varepsilon^{183}W_{meas.}$ (± 2 s.e.) | $\varepsilon^{183}W_{corr.}$* (± 2σ) | $\varepsilon^{182}W_{meas.}$ (± 2 s.e.) | $\varepsilon^{182}W_{corr.}$* (± 2σ) | $\varepsilon^{184}W_{meas.}$ (± 2 s.e.) | $\varepsilon^{184}W_{corr.}$* (± 2σ) |
|---|---|---|---|---|---|---|---|---|---|
| | | | | normalized to $^{186}W/^{184}W$ = 0.92767 | | | normalized to $^{186}W/^{183}W$ = 1.98590 | | |
| N9C06.1 | 0 | 702 | -0.03 ± 0.08 | -0.17 ± 0.08 | -0.02 ± 0.10 | 0.20 ± 0.08 | -0.03 ± 0.13 | 0.12 ± 0.05 | 0.02 ± 0.06 |
| N9C06.2 | – | – | -0.02 ± 0.08 | -0.16 ± 0.07 | -0.01 ± 0.09 | 0.20 ± 0.07 | -0.02 ± 0.12 | 0.11 ± 0.05 | 0.01 ± 0.06 |
| N9C06.3 | – | – | 0.07 ± 0.10 | -0.17 ± 0.08 | 0.05 ± 0.10 | 0.30 ± 0.08 | 0.07 ± 0.13 | 0.11 ± 0.05 | -0.04 ± 0.07 |
| N9C06.4 | – | – | 0.01 ± 0.08 | -0.16 ± 0.07 | 0.01 ± 0.09 | 0.23 ± 0.08 | 0.01 ± 0.12 | 0.11 ± 0.05 | -0.01 ± 0.06 |
| N9C06.5 | – | – | 0.02 ± 0.09 | -0.19 ± 0.08 | 0.01 ± 0.10 | 0.27 ± 0.09 | 0.02 ± 0.13 | 0.13 ± 0.05 | -0.01 ± 0.07 |
| N9C06.6 | – | – | 0.15 ± 0.09 | -0.06 ± 0.07 | 0.13 ± 0.09 | 0.25 ± 0.07 | 0.17 ± 0.12 | 0.04 ± 0.05 | -0.08 ± 0.06 |
| N9C07.1 | – | – | 0.01 ± 0.09 | -0.13 ± 0.07 | 0.00 ± 0.09 | 0.17 ± 0.08 | 0.00 ± 0.13 | 0.09 ± 0.05 | 0.00 ± 0.06 |
| N9C07.2 | – | – | 0.06 ± 0.08 | -0.08 ± 0.07 | 0.05 ± 0.10 | 0.17 ± 0.08 | 0.06 ± 0.13 | 0.06 ± 0.05 | -0.03 ± 0.06 |
| N9C07.3 | – | – | -0.02 ± 0.08 | -0.12 ± 0.07 | -0.02 ± 0.09 | 0.14 ± 0.07 | -0.02 ± 0.12 | 0.08 ± 0.05 | 0.01 ± 0.06 |
| N9C07.4 | – | – | 0.07 ± 0.09 | -0.12 ± 0.08 | 0.05 ± 0.10 | 0.22 ± 0.07 | 0.06 ± 0.13 | 0.08 ± 0.05 | -0.03 ± 0.06 |
| N9C07.5 | – | – | 0.06 ± 0.09 | -0.12 ± 0.07 | 0.07 ± 0.09 | 0.25 ± 0.07 | 0.10 ± 0.12 | 0.08 ± 0.05 | -0.05 ± 0.06 |
| N9C07.6 | – | – | 0.06 ± 0.10 | -0.13 ± 0.08 | 0.06 ± 0.10 | 0.25 ± 0.08 | 0.08 ± 0.13 | 0.09 ± 0.05 | -0.04 ± 0.06 |
| N | – | – | 12 | 12 | 12 | 12 | 12 | 12 | 12 |
| Mean | – | – | 0.04 | -0.14 | 0.03 | 0.22 | 0.04 | 0.09 | -0.02 |
| 2 s.d. | – | – | 0.10 | 0.08 | 0.09 | 0.09 | 0.12 | 0.05 | 0.06 |
| 95% CI | – | – | 0.03 | 0.02 | 0.03 | 0.03 | 0.04 | 0.02 | 0.02 |

Different numbers (22–30, 06–07) denote separate digestions of ~0.5 g (BHVO-2) or ~0.3 g (SRM 129c) standard material, which were processed through the full chemical separation procedure and analyzed with each set of samples. Each line represents a single analysis that consumed ~25 ng of W (run at ~30 ng/ml).

*Internally corrected for a small analytical $^{183}W$ effect after Budde et al. (2022); all uncertainties propagated.

**Table S2:** Titanium isotope data for the geological reference materials.

JB-2 (GSJ), Japanese basalt

| ID | $\varepsilon^{46}$Ti | $\varepsilon^{48}$Ti | $\varepsilon^{50}$Ti |
|---|---|---|---|
| | normalized to $^{49}$Ti/$^{47}$Ti = 0.749766 | | |
| JB-2C04.1 | -0.13 | 0.03 | 0.05 |
| JB-2C04.2 | -0.10 | 0.08 | 0.00 |
| JB-2C04.3 | -0.01 | -0.03 | -0.07 |
| JB-2C04.4 | 0.05 | 0.14 | 0.26 |
| JB-2C04.5 | -0.09 | -0.06 | 0.02 |
| JB-2C04.6 | -0.21 | -0.04 | 0.01 |
| JB-2C04.7 | 0.00 | 0.04 | -0.12 |
| JB-2C04.8 | -0.02 | 0.03 | -0.04 |
| JB-2C04.9 | -0.02 | 0.07 | 0.19 |
| JB-2C04.10 | 0.05 | -0.08 | -0.24 |
| JB-2C04.11 | 0.08 | 0.04 | -0.11 |
| JB-2C04.12 | -0.12 | 0.03 | -0.04 |
| JB-2C04.13 | 0.04 | -0.05 | 0.04 |
| JB-2C04.14 | -0.01 | 0.26 | 0.28 |
| JB-2C04.15 | -0.07 | 0.04 | 0.13 |
| JB-2C04.16 | 0.19 | 0.14 | 0.03 |
| JB-2C04.17 | -0.05 | -0.18 | -0.02 |
| JB-2C04.18 | -0.06 | -0.13 | -0.01 |
| JB-2C04.19 | -0.09 | 0.04 | -0.04 |
| JB-2C04.20 | -0.29 | -0.05 | 0.06 |
| JB-2C04.21 | 0.18 | -0.11 | -0.02 |
| JB-2C04.22 | 0.15 | 0.04 | -0.04 |
| JB-2C04.23 | 0.30 | 0.01 | 0.05 |
| JB-2C04.24 | 0.00 | -0.20 | -0.25 |
| JB-2C04.25 | -0.19 | -0.11 | 0.10 |
| N | 25 | 25 | 25 |
| Mean | -0.02 | 0.00 | 0.01 |
| 2 s.d. | 0.27 | 0.21 | 0.25 |
| 95% CI | 0.05 | 0.04 | 0.05 |

In total, ~0.5 g of JB-2 were digested and a solution aliquot equivalent to ~30 μg Ti was processed through the full chemical separation procedure and analyzed along with the samples. Each line represents a single analysis that consumed ~70 ng of Ti (run at ~200 ng/ml).

**Table S3:** Summary of ages for CB, CH and CR chondrites using different chronometers.

| Method | Group | Samples | Anchor | Absolute age [Ma] | (± 2σ) | Relative age [Ma] | (± 2σ) | Refrence |
|---|---|---|---|---|---|---|---|---|
| **Hf-W** | CB | Gujba (CBa) metal | CAI | - | - | 4.2 | 1.8 | this study, model age |
| | | Bencubbin (CBa) metal | CAI | - | - | 3.3 | 2.0 | this study, model age |
| | | combined CB metal (wt. av., $n$ = 2) | CAI | - | - | 3.8 | 1.3 | this study, model age |
| | $CB_b$/CH | Isheyevo metal | CAI | - | - | 3.7 | 1.7 | this study, model age |
| | CH | Acfer 182 isochron | CAI | - | - | 3.8 | 1.2 | this study, isochron |
| | | Acfer 182 metal (M-1) | CAI | - | - | 4.2 | 2.2 | this study, model age |
| | | Acfer 182 metal (M-2) | CAI | - | - | 3.5 | 2.1 | this study, model age |
| | | combined CH metal (wt. av., $n$ = 2) | CAI | - | - | 3.8 | 1.5 | this study, model age |
| | CR | combined CR isochron | CAI | - | - | 3.6 | 0.6 | Budde et al. (2018) |
| | | CR chondrule isochron | CAI | - | - | 3.7 | 0.8 | Budde et al. (2018) |
| **Mn-Cr** | CB | Gujba (CBa) chondrule/metal | D'Orbigny | - | - | 4.0 | 0.9 | Yamashita et al. (2010) |
| **I-Xe** | CB | Gujba (CBa) chondrule | Shallowater | - | - | 4.1 | 1.3 | Gilmour et al. (2009)[a] |
| | | HaH 237 chondrule | Shallowater | - | - | 5.2 | 0.3 | Pravdivtseva et al. (2017) |
| **Al-Mg** | CB | HH237 & Gujba chondrules | canonical | - | - | >3.5 | - | Olsen et al. (2013) |
| | | HaH 237 & QUE94411 chondrules | canonical | - | - | >3.5 | - | Gounelle et al. (2007) |
| | CH | Acfer 182/214 chondrules | canonical | - | - | >3.5 | - | Krot et al. (2017)[b] |
| | CR | Bulk CR chondrules (wt. mean) | canonical | - | - | 3.8 | 0.2 | Schrader et al. (2017); Nagashima et al. (2014) |
| **Pb-Pb** | CB | Gujba (Cba) chondrules | - | 4562.5 | 0.2 | 4.8 | 0.3 | Bollard et al. (2015) |
| | | Gujba (CBa) chondrules | - | 4561.7 | 0.5 | 5.6 | 0.5 | Krot et al. (2005)[a] |
| | | HH237 (CBb) chondrules | - | 4561.8 | 0.9 | 5.5 | 0.9 | Krot et al. (2005)[a] |
| | CR | Bulk CR chondrules (wt. mean) | - | 4563.6 | 0.6 | 3.7 | 0.6 | Amelin et al. (2002) |

[a] Ages corrected for erroneously assumed $^{238}U/^{235}U$ = 137.88 using appropriate $^{238}U/^{235}U$, respectively. See main text for explanation.
[b] The vast majority of investigated samples is characterized by $(^{26}Al/^{27}Al)_0$ ratios that refer to formation ages of at least 3.5 Ma after CAI formation. Only very few samples show evidence for older formation ages (e.g., Krot et al., 2017).

# 12 References in Supplementary Material